\documentclass[a4paper,11pt]{article}
\usepackage[english]{babel}
\usepackage{jheppub}
\pdfoutput=1
\usepackage[table]{xcolor}
\usepackage[T1]{fontenc}
\usepackage{bbold}
\usepackage{float}
\usepackage{amsmath}
\usepackage{empheq}
\usepackage{booktabs}
\usepackage{color}
\usepackage[utf8]{inputenc}
\usepackage{xspace}
\usepackage{scalerel}
\usepackage[most]{tcolorbox}
\usepackage{multirow}
\usepackage{scalefnt}
\usepackage{bold-extra}
\usepackage[shortlabels]{enumitem}
\usepackage[tikz]{bclogo}
\usepackage{subcaption}
\usepackage{cancel}
\usepackage{verbatim}
\usetikzlibrary{arrows,shapes}
\usepackage{afterpage}
\usepackage{graphicx,grffile}
\usepackage{xspace}
\usepackage[normalem]{ulem}
\usepackage{makecell}
\usepackage{tabularx,colortbl}
\usepackage{nicematrix}

\newcommand*{\olr}{\overleftrightarrow}
\newcommand*{\olarrow}{\overleftarrow}

\newcommand{\br}[1]{{\mathrm{BR}(#1)}}

\usepackage{xcolor}

\newtcolorbox{empheqboxed}{colback=white!35,
 colframe=black,
 width=\textwidth,
 sharpish corners,
 top=-2mm, 
 bottom=0pt
}

\title{SMEFT Restrictions On Exclusive $b \to u \ell \nu$ Decays}

\author[a]{Admir Greljo,}
\author[a]{Jakub Salko,}
\author[a,b]{Aleks Smolkovi\v c,}
\author[c]{Peter Stangl}

\emailAdd{admir.greljo@unibas.ch}
\emailAdd{j.salko@unibas.ch}
\emailAdd{smolkovic@itp.unibe.ch}
\emailAdd{peter.stangl@cern.ch}

\affiliation[a]{Department of Physics, University of Basel, Klingelbergstrasse 82,  CH-4056 Basel, Switzerland}

\affiliation[b]{Albert Einstein Center for Fundamental Physics, Institut f\"{u}r Theoretische Physik, Universit\"{a}t Bern, Sidlerstrasse 5, CH-3012 Bern, Switzerland.}

\affiliation[c]{CERN, Theoretical Physics Department, CH-1211 Geneva 23, Switzerland}

\abstract{
Exclusive semileptonic $b$ hadron decays ($b \to u \ell \nu$) serve as a sandbox for probing strong and electroweak interactions and for extracting the CKM element $V_{ub}$. Instead, this work investigates their underexplored potential to reveal new short-distance physics. Utilizing SMEFT as a conduit to chart territory beyond the SM, we demonstrate that substantive new physics contributions in $b \to u \ell \nu$ are necessarily linked to correlated effects in rare neutral-current $b$ decays, neutral $B$ meson mixing or high-mass Drell-Yan tails. We find that measurements of the latter processes strongly restrict the allowed deviations in the former. A complete set of tree-level mediators, originating from a perturbative ultraviolet model and matching at dimension 6, is thoroughly explored to support this assertion. As a showcase application, we examine the feasibility of a new physics interpretation of the recent tension in exclusive $|V_{ub}|$ extraction from $B \to V \ell \nu$ where $V=(\rho,\omega)$.
}


\preprint{CERN-TH-2023-111}

\begin{document}

\maketitle

\newpage

\section{Introduction}
\label{sec:intro}

The study of $b$ hadron decays has attracted increasing attention in recent years and promises to remain an active research area in this decade. The LHCb experiment, presently underway at the Large Hadron Collider (LHC) at CERN, stands as a pivotal complement to the B-factory studies conducted at the late BaBar and Belle experiments. As we advance into the decade, the Belle II experiment at Super KEKB is rapidly gaining ground, with its luminosity nearing that of the BaBar dataset. Forecasts suggest that Belle II will outstrip its predecessor, the original Belle experiment, in the forthcoming years. The projected data sets from Belle~II~\cite{Belle-II:2018jsg} and LHCb~\cite{LHCb:2018roe} have the potential to reshape our understanding of $b$~physics.

The theoretical framework of weak decays is constructed on the foundation of effective field theory, which advocates the factorization of long- and short-distance contributions. Long-distance contributions are encapsulated by form factors, a domain wherein recent advancements in lattice QCD have been particularly substantial~\cite{FlavourLatticeAveragingGroupFLAG:2021npn, DiCanto:2022icc, Dalgic:2006dt, Lattice:2015tia, Flynn:2015mha, Colquhoun:2022atw}. Short-distance contributions, conversely, are expressed via the Wilson coefficients (WCs) of the Weak Effective Theory (WET)~\cite{Buchalla:1995vs, Buras:2020xsm, Jenkins:2017jig, Jenkins:2017dyc} and can be calculated perturbatively in the Standard Model (SM). Any deviation in the WCs from their SM predictions would flag the presence of short-distance new physics (NP). When the NP scale surpasses the electroweak (EW) scale, an encompassing model-independent interpretation can be depicted through the Standard Model Effective Field Theory (SMEFT)~\cite{Buchmuller:1985jz, Grzadkowski:2010es, Brivio:2017vri, Isidori:2023pyp, Giudice:2007fh, Henning:2014wua}. 

This succession of effective theories offers a methodical blueprint to describe the variety of possible short-distance physics beyond the SM (BSM). By hypothesizing a weakly-coupled ultraviolet (UV) theory as the subsequent layer of physics, we can facilitate perturbative matching calculations to the SMEFT~\cite{deBlas:2017xtg, Fuentes-Martin:2020udw, Cohen:2020qvb, Carmona:2021xtq}. The renormalization group (RG) evolution from the NP scale down to the hadron scale~\cite{Alonso:2013hga, Jenkins:2013wua, Jenkins:2013zja, Aebischer:2017gaw, Jenkins:2017dyc}, encompassing an intermediary SMEFT to WET matching~\cite{Aebischer:2015fzz,Jenkins:2017jig,Dekens:2019ept}, yields reliable predictions. This framework permits the systematic organization of BSM effects, which may be observable in a given weak hadron decay at each order in the EFT and loop expansion. In practice, our interest is often focused on the few leading  orders which are capable of producing a substantial effect. Such a comprehensive classification study for a given weak decay enables us to identify other observables correlated at both low and high energies, thereby proposing a robust test for NP.

In accordance with this perspective, the principal objective of this paper is to explore potential BSM effects in exclusive $b$ hadron decays, particularly those undergoing the quark-level transition $b \to u \ell \nu$. Within the SM, these transitions are classified as tree-level weak decays, facilitated through the exchange of a $W$ boson, and regulated by the Cabibbo-Kobayashi-Maskawa (CKM) matrix element $V_{ub}$. Traditionally, these decays serve as a sandbox for probing strong and EW interactions and for extracting the absolute value of the CKM element $|V_{ub}|$. In the course of this study, we venture into the relatively unexamined potential of these decays to reveal insights into new short-distance physics.
While model-independent constraints on new physics from $b \to u \ell \nu$ processes have been studied in the WET~\cite{Sahoo:2017bdx, Banelli:2018fnx, Colangelo:2019axi, Fleischer:2021yjo, Leljak:2023gna}, employing the SMEFT allows us to consider crucial correlations imposed by phenomena such as flavor-changing neutral currents, along with other observables like the tails of high-mass Drell-Yan distributions.

The extraction of the CKM matrix elements $|V_{qb}|$ (where $q=u,c$) has exhibited slight discrepancies between determinations derived from exclusive and inclusive decays~\cite{HeavyFlavorAveragingGroup:2022wzx, Ricciardi:2019zph, Ricciardi:2021shl, Bordone:2021oof, Capdevila:2021vkf, Bernlochner:2022ucr}, and the NP explanation was found to be challenging~\cite{Bernlochner:2012bc, Enomoto:2014cta, Crivellin:2014zpa, Bernlochner:2014ova, Colangelo:2016ymy}. A recent reconsideration of the $B \to \pi$ form factors~\cite{Leljak:2021vte} reveals a $|V_{ub}|$ value congruent with the most recent inclusive determination from Belle~\cite{Belle:2021eni}; see also~\cite{Belle:2023asx}. While this puzzle seems to be settled, a tension remains in the $|V_{ub}|$ determination from the exclusive $B \to \{\rho,\omega\} \ell \nu$ decays~\cite{Bernlochner:2021rel}. As an ancillary application of our study, we offer insights into the feasibility of a NP interpretation. 

The inclusive $B \to X_u \ell \nu$ decays have sizable uncertainties, partially attributed to the subtraction of large $B \to X_c \ell \nu$ backgrounds. Furthermore, the requisite background suppression cuts challenge the theoretical description based on the heavy quark expansion~\cite{Lange:2005yw, Gambino:2007rp, Andersen:2005mj}. Consequently, the inclusive decays will not be further addressed, and the scope of our work is limited to the exclusive $b \to u \ell \nu$ decays.

The structure of this paper is organized as follows: In Section~\ref{sec:WET}, we undertake a comprehensive analysis of $b \to u \ell \nu$ decays within the framework of the WET. Progressing to Section~\ref{sec:SMEFT}, we transition to the SMEFT and execute a global analysis that also takes into consideration correlations with other data sets. Subsequently, in Section~\ref{sec:models}, we enumerate all tree-level mediator models that match onto the SMEFT scenarios and probe the implied correlations in detail. The paper reaches its conclusion in Section~\ref{sec:conc}. Furthermore, Appendix~\ref{app:Vub} provides additional details on the determination of $|V_{ub}|$ in the WET, while Appendix~\ref{app:mediators} delves into tree-level models.

\section{WET analysis of $b \to u \ell \nu$ decays}
\label{sec:WET}

This section presents the theoretical framework, the WET, to describe the effects of NP at short distances in $b \to u \ell \nu$ decays (Subsection~\ref{sec:WET-setup}). We detail the relevant set of operators and their contributions to observables. Furthermore, we discuss the implementation of the experimental data and theoretical predictions within the \texttt{flavio} framework~\cite{Straub:2018kue}. Finally, in Subsection~\ref{sec:WET-fits}, we offer a comprehensive interpretation of the data in the context of the WET. This serves as the starting point for the SMEFT analysis in Section~\ref{sec:SMEFT}.

\subsection{Setup}
\label{sec:WET-setup}

Here we focus on fully leptonic and semileptonic exclusive $B$ meson decays with the underlying $b\to u l \nu$ transition. We employ the following weak effective Hamiltonian,
\begin{equation}
\mathcal{H}_\mathrm{eff} = \mathcal{H}_\mathrm{eff}^\mathrm{SM} + \frac{4 G_F}{\sqrt{2}} V_{ub} \sum_{i, l} C_i^{(l)} O_i^{(l)}+ \mathrm{h.c.}\,,
\label{eq:WET}
\end{equation}
where $O_i^{(l)}$ are local effective operators, $C_i^{(l)}$ are WCs encoding contributions of short-distance NP, $V_{ub}$ is the CKM matrix element, $G_F$ is the Fermi constant, and $l$ represents the lepton flavor ($l=e, \mu, \tau$).\footnote{Throughout this work we assume lepton flavor conservation.} We consider the following set of local operators at mass dimension $6$:
\begin{align}
O_{V_L}^{(l)} &=
(\bar{u}_L \gamma^{\mu} b_L)(\bar{l}_L \gamma_\mu \nu_{l L})\,,
&
O_{V_R}^{(l)} &=
(\bar{u}_R \gamma^{\mu} b_R)(\bar{l}_L \gamma_\mu \nu_{l L})\,,\label{eq:OVLOVR}
\\
O_{S_L}^{(l)} &=
(\bar{u}_R b_L)(\bar{l}_R \nu_{l L})\,,
&
O_{S_R}^{(l)} &=
(\bar{u}_L b_R)(\bar{l}_R \nu_{l L})\,,\label{eq:OSLOSR}
\\
O_{T}^{(l)} &=
(\bar{u}_R \sigma^{\mu\nu} b_L)(\bar{l}_R \sigma_{\mu\nu} \nu_{l L})\,.\label{eq:OT}
\end{align}
In the SM, only the left-handed vector operator is generated through a tree-level exchange of the $W$ boson, leading to $C_{V_L}^{(l) \mathrm{SM}} = 1$ using the same normalization as in Eq.~\eqref{eq:WET}.\footnote{The EW corrections are included as $C_{V_L}^{(l) \mathrm{SM}} = 1 + \frac{\alpha_e}{\pi}\log\left(\frac{m_Z}{\mu_b}\right)$ with $\mu_b=4.8\ \text{GeV}$.} On the other hand, short-distance NP can, in general, generate the full set of operators in Eqs.~\eqref{eq:OVLOVR} - \eqref{eq:OT}. Various low-energy observables, such as the branching ratios of leptonic and semileptonic exclusive $B$ decays, are highly sensitive probes of various combinations of the corresponding WCs, as implied by Lorentz symmetry and invariance of QCD under parity.

The fully leptonic decays $B \to l \nu_l$ are sensitive to axial, $C_A^{(l)} \equiv C_{V_R}^{(l)} - C_{V_L}^{(l)}$, and pseudoscalar, $C_P^{(l)} \equiv C_{S_R}^{(l)} - C_{S_L}^{(l)}$, combinations of the WCs, via
\begin{equation}
\label{eq:Blnu}
    \frac{\br{B\to l \nu_l}}{\br{B\to l \nu_l}_\mathrm{SM}} = \left|1-(C_{V_R}^{(l)} - C_{V_L}^{(l)}) + \frac{m_B^2}{m_l (m_b+m_u)} (C_{S_R}^{(l)} - C_{S_L}^{(l)})\right|^2\,.
\end{equation}
Note that the pseudoscalar contribution to the branching ratios of the leptonic $B$ decays is helicity enhanced compared to the axial contribution, rendering them highly efficient probes of the pseudoscalar operator.

The semileptonic decay modes of $B$ mesons into a pseudoscalar meson $P$ are sensitive to vectorial, $C_V^{(l)} \equiv C_{V_R}^{(l)} + C_{V_L}^{(l)}$, scalar $C_S^{(l)} \equiv C_{S_R}^{(l)} + C_{S_L}^{(l)}$, and tensor $C_{T}^{(l)}$ WCs. The differential decay width (relative to the SM one) can be written as~\cite{Tanaka:2012nw, Sakaki:2013bfa}

\begin{equation}
\label{eq:BPlnu}
   \begin{split}
      \frac{d\Gamma(B \to P l \nu)/dq^2} {d\Gamma(B \to P l \nu)^\mathrm{SM} / dq^2}=& 
      \left|1 + (C_{V_R}^{(l)} + C_{V_L}^{(l)})\right|^2 \left[ \left( 1 + \frac{m_l^2}{2q^2} \right) H_{V,0}^{s\,2} + \frac{3}{2}\frac{m_l^2}{q^2} \, H_{V,t}^{s\,2} \right] \\
      &+ \frac{3}{2} |C_{S_R}^{(l)} + C_{S_L}^{(l)}|^2 \, H_S^{s\,2} + 8|C_T^l|^2 \left( 1+ \frac{2m_l^2}{q^2} \right) \, H_T^{s\,2} \\
      &+ 3\mathrm{Re}[ ( 1 + (C_{V_R}^{(l)} + C_{V_L}^{(l)}) ) (C_{S_R}^{(l)*} + C_{S_L}^{(l)*}) ] \frac{m_l}{\sqrt{q^2}} \, H_S^s H_{V,t}^s \\
      &- 12\mathrm{Re}[ ( 1 + (C_{V_R}^{(l)} + C_{V_L}^{(l)})) C_T^{(l)*} ] \frac{m_l}{\sqrt{q^2}} \, H_T^s H_{V,0}^s \biggl.\,,
   \end{split}
\end{equation}
where $q^2$ is the momentum transfer squared, and $H_{V,0}^{s\,2}, H_{V,t}^{s\,2}, H_S^{s\,2}, H_T^{s\,2}$ are hadronic matrix elements, parameterized by the three hadronic form factors $f_{+,0,T}(q^2)$ (see Ref.~\cite{Sakaki:2013bfa} for explicit expressions). We use the latest determination of the $B\to \pi$ form factors from Ref.~\cite{Leljak:2021vte} where a combined fit to light-cone sum rules (LCSR) and lattice QCD data was performed.

The semileptonic decay modes of $B$ mesons into vector meson $V$ are rich in structure, with the differential decay width (relative to the SM one) given as~\cite{Tanaka:2012nw, Sakaki:2013bfa}
\begin{equation}
\label{eq:BVlnu}
   \begin{split}
    \frac{d\Gamma(B \to V l \nu)/dq^2} {d\Gamma(B \to V l \nu)^\mathrm{SM} / dq^2}  = & \left( |1+ C_{V_L}^{(l)}|^2 + |C_{V_R}^{(l)}|^2 \right) \left[ \left( 1 + \frac{m_l^2}{2q^2} \right) \left( H_{V,+}^2 + H_{V,-}^2 + H_{V,0}^2 \right) + \frac{3}{2}\frac{m_l^2}{q^2} \, H_{V,t}^2 \right] \\
      & - 2\mathrm{Re}[(1 + C_{V_L}^{(l)}) C_{V_R}^{(l)*}] \left[ \left( 1 + \frac{m_l^2}{2q^2} \right) \left( H_{V,0}^2 + 2 H_{V,+} H_{V,-} \right) + \frac{3}{2}\frac{m_l^2}{q^2} \, H_{V,t}^2 \right] \\
      &  + \frac{3}{2} |C_{S_R}^{(l)} - C_{S_L}^{(l)}|^2 \, H_S^2 + 8|C_T^l|^2 \left( 1+ \frac{2m_l^2}{q^2} \right) \left( H_{T,+}^2 + H_{T,-}^2 + H_{T,0}^2  \right) \\
      &  + 3\mathrm{Re}[ ( 1 - (C_{V_R}^{(l)} - C_{V_L}^{(l)})) (C_{S_R}^{(l)*} - C_{S_L}^{(l)*})  ] \frac{m_l}{\sqrt{q^2}} \, H_S H_{V,t} \\
      & - 12\mathrm{Re}[ (1 + C_{V_L}^{(l)}) C_T^{(l)*} ] \frac{m_l}{\sqrt{q^2}} \left( H_{T,0} H_{V,0} + H_{T,+} H_{V,+} - H_{T,-} H_{V,-} \right) \\
      &  + 12\mathrm{Re}[ C_{V_R}^{(l)} C_T^{(l)*} ] \frac{m_l}{\sqrt{q^2}} \left( H_{T,0} H_{V,0} + H_{T,+} H_{V,-} - H_{T,-} H_{V,+} \right) \,,
   \end{split}
\end{equation}

where again, the explicit expressions of the hadronic matrix elements can be found in Ref.~\cite{Sakaki:2013bfa}, parameterized in terms of the form factors $A_{0,1,2}(q^2)$, $V(q^2)$ and $T_{1,2,3}(q^2)$. The form factors for the $B\to \omega$ and $B\to \rho$ transitions have been determined only by the light-cone sum rules~\cite{Bharucha:2015bzk}.

In the following, we contrast the WET with the available experimental data on the leptonic and semileptonic $b\to u l \nu$ decays, focusing mostly on channels with light leptons. We assume exact isospin symmetry, and the spectra are to be understood as CP-averaged. Currently, the best determination of $\br{B\to e \nu}$ and $\br{B\to \mu \nu}$ comes from Belle~\cite{Belle:2006tbq, Belle:2019iji}, whereas for $\br{B\to \tau \nu}$ we use the latest PDG average~\cite{ParticleDataGroup:2022pth} of Belle~\cite{Belle:2015odw,Belle:2012egh} and BaBar~\cite{BaBar:2012nus,BaBar:2009wmt} measurements,
\begin{equation}
\label{eq:Blnu-exp}
\begin{split}
    \br{B\to e \nu}^\mathrm{exp} &< 9.8\times 10^{-7} ~\mathrm{at}~90\% ~\mathrm{CL} \,,\\
    \br{B\to \mu \nu}^\mathrm{exp} &= (5.3 \pm 2.0 \pm 0.9) \times 10^{-7} \,,\\
    \br{B\to \tau \nu}^\mathrm{exp} &= (1.09 \pm 0.24)\times 10^{-4} \,.\\
\end{split}
\end{equation}
It is important to note that the obtained results are based on the assumption that backgrounds from semileptonic decays are NP-free. This assumption adds complexity to the global fit process, which will be discussed in detail later in this article.

As for the semileptonic transitions, we use the latest available data on the differential branching ratios of $B\to \pi l \nu$, $B\to \omega l \nu$ and $B\to \rho l \nu$ from Belle~\cite{Belle:2010hep,Belle:2013hlo} and BaBar~\cite{BaBar:2010efp,BaBar:2012thb, BaBar:2012dvs}, as combined in the latest HFLAV averages~\cite{Bernlochner:2021rel, HeavyFlavorAveragingGroup:2022wzx}. Unfortunately, the data is only reported as a combination of the electron and muon channels, hence for the majority of the following discussion (unless stated otherwise), we will assume lepton flavor universality (LFU) in light leptons, writing $\ell \equiv e, \mu$. From the Belle analysis of $B\to D\ell \nu$ \cite{Belle:2015pkj}, we expect a similar experimental sensitivity to electrons and muons in the decay modes discussed here. Hence we assume $C^{(\ell)} = 1/2 (C^{(e)} + C^{(\mu)})$. There is also an upper limit on $B\to \pi \tau \nu$ from Belle~\cite{Belle:2015qal}, which does not play a significant role in the scenarios studied here at the current level of experimental precision.

The predictions and measurements discussed above are available in the open source python package \texttt{flavio}~\cite{Straub:2018kue} and part of the global SMEFT likelihood \texttt{smelli}~\cite{Aebischer:2018iyb,Stangl:2020lbh}. As part of performing the study discussed here, we have updated the measurements of $B\to \mu \nu$, $B\to \pi \ell \nu$, $B \to \omega \ell \nu$ and $B\to \rho \ell \nu$ in the database of measurements. In our numerical analysis, we take into account the new physics dependence of the theory uncertainties and their correlations, as described in Ref.~\cite{Altmannshofer:2021qrr}.

\subsection{Results}
\label{sec:WET-fits}

Here we perform a comprehensive analysis of the constraining power of the above processes on the model-independent parameterization of short-distance NP effects by means of the WET Hamiltonian in Eq.~\eqref{eq:WET}. In the main results of this paper, we fix $|V_{ub}|$ in Eq.~\eqref{eq:WET} to $|V_{ub}| = 3.73\times 10^{-3}$, a value which is compatible with the recent global fits of the CKM matrix parameters~\cite{HeavyFlavorAveragingGroup:2022wzx}. We dedicate Appendix \ref{app:Vub} to the issue of extracting $|V_{ub}|$ in an EFT setting.

\begin{figure}[t]
     \centering
     \begin{subfigure}[b]{0.49\textwidth}
         \centering
         \includegraphics[width=\textwidth]{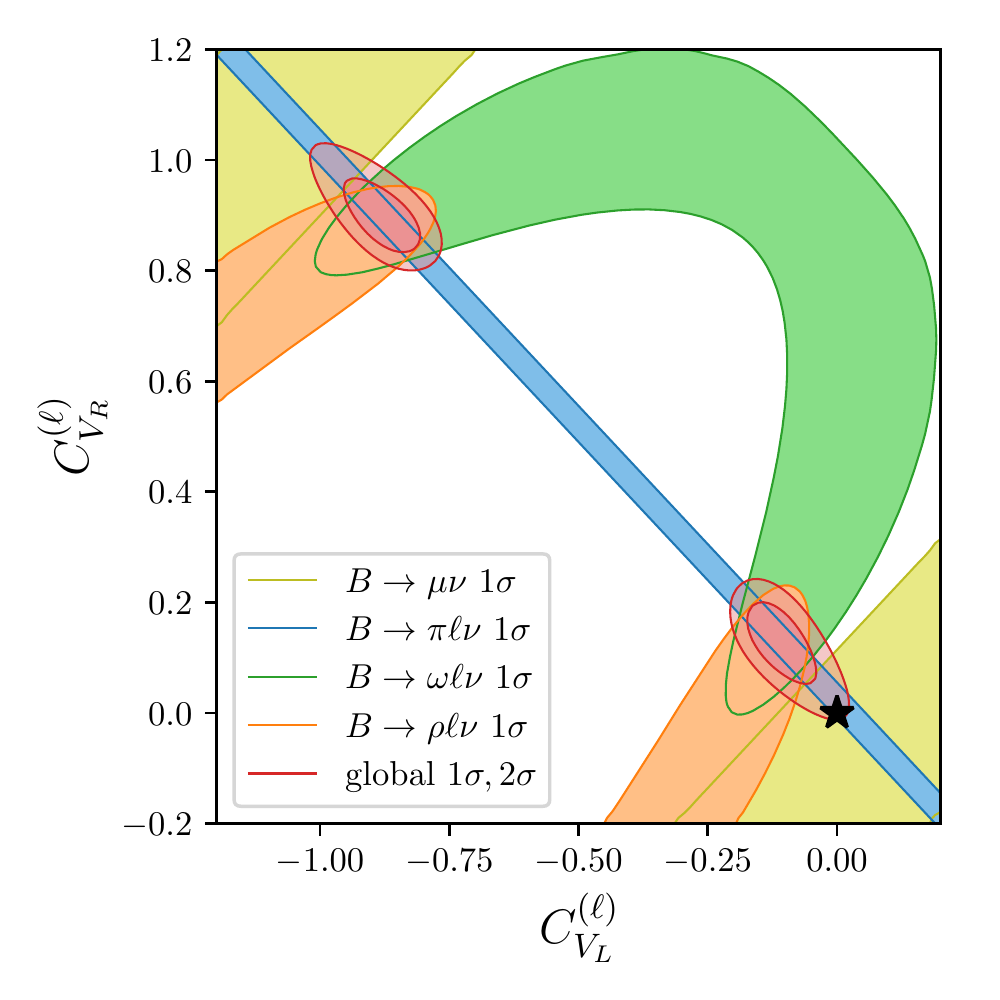}
     \end{subfigure}~
     \begin{subfigure}[b]{0.49\textwidth}
         \centering
         \includegraphics[width=\textwidth]{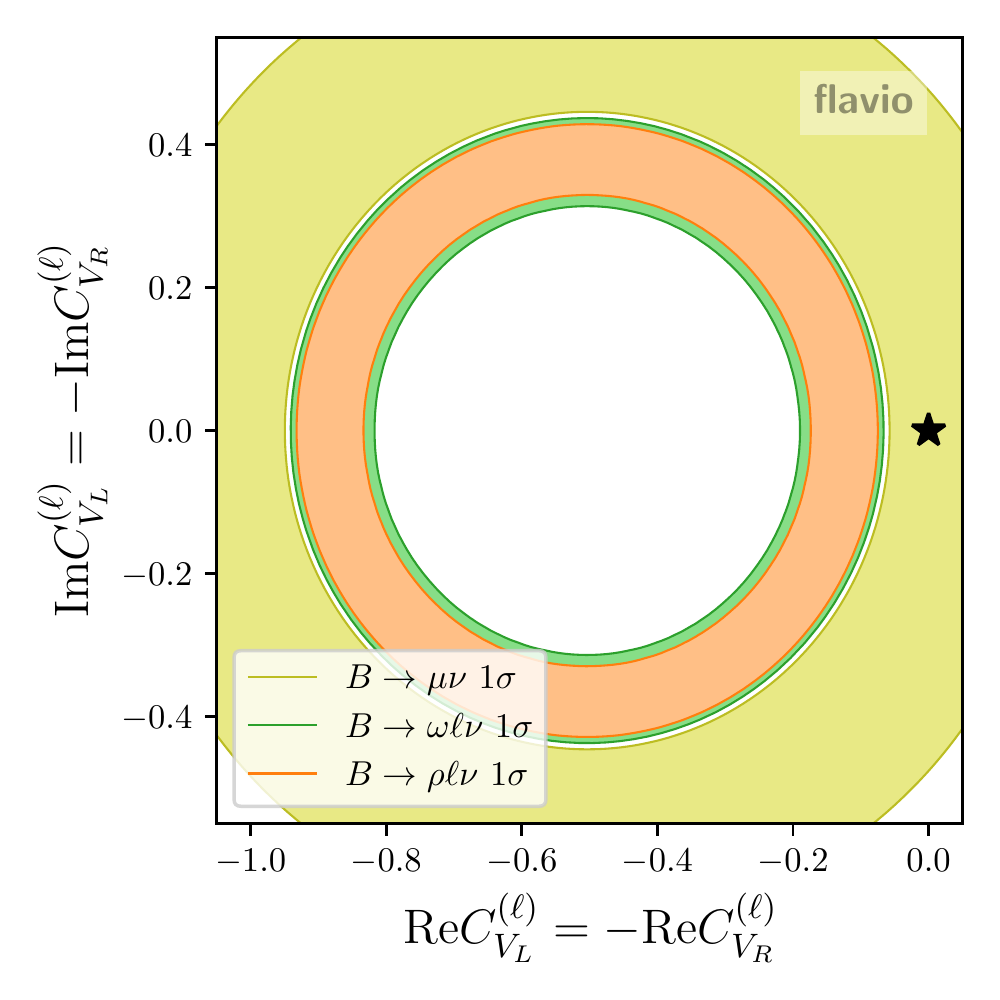}
     \end{subfigure}\\
     \begin{subfigure}[b]{0.49\textwidth}
         \centering
         \includegraphics[width=\textwidth]{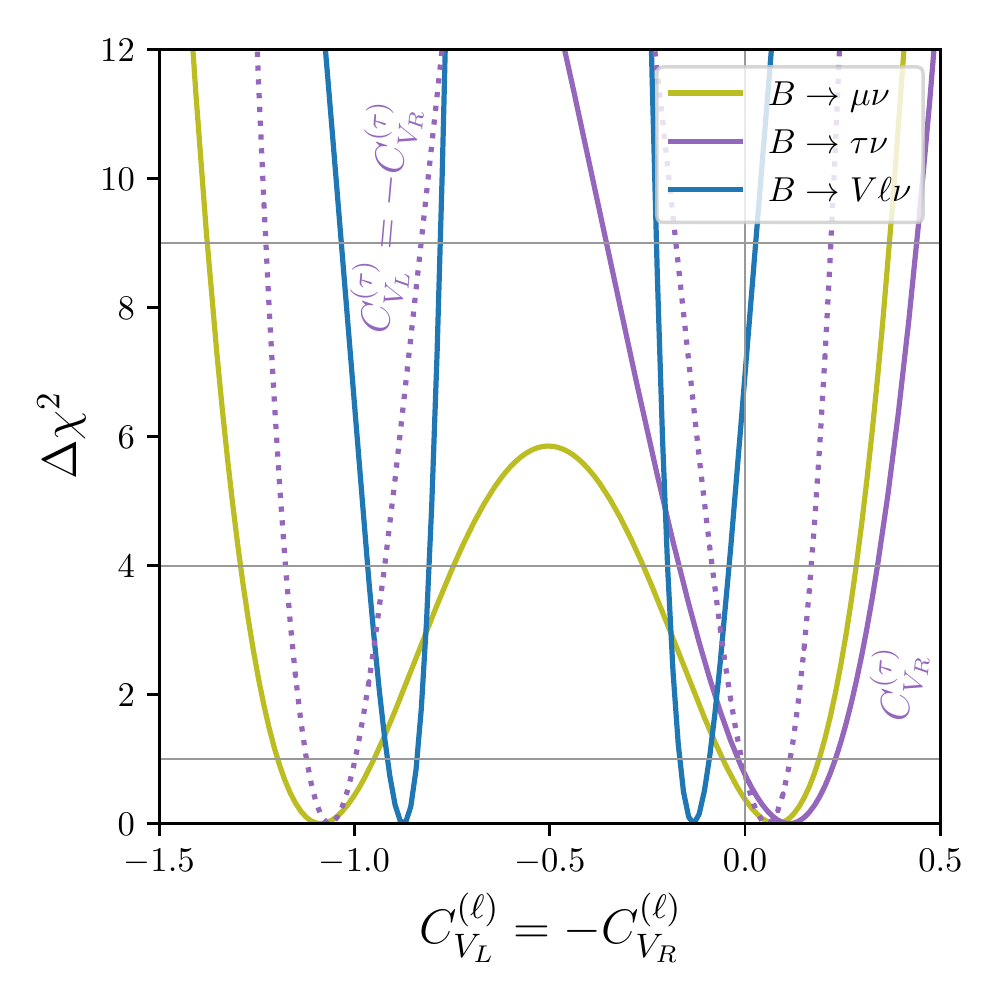}
     \end{subfigure}
        \caption{
        \textbf{Left:} 2D contours in the scenario $(C_{V_L}^{(\ell)}, C_{V_R}^{(\ell)})$, assuming LFU for light leptons.
        \textbf{Right:} 2D contours in the complex plane of the scenario $(C_{V_L}^{(\ell)}=-C_{V_R}^{(\ell)})$, assuming LFU for light leptons.
        \textbf{Bottom:} $\Delta \chi^2$ as a function of $C_{V_L}^{(\ell)}=-C_{V_R}^{(\ell)}$, assuming LFU for light leptons, and either LFU in $C_{V_R}^{(\tau)}$ or $C_{V_L}^{(\tau)}=-C_{V_R}^{(\tau)}$. Unless denoted otherwise (as in the right plot), we assume the WCs to be real and drop the symbol $\mathrm{Re}$ for brevity. See Section~\ref{sec:WET-fits} for details.
        }
        \label{fig:wet_CVLCVR}
\end{figure}
Firstly, we focus on the left- and right-handed vector operators in Eq.~\eqref{eq:OVLOVR}. On the upper left plot in Figure~\ref{fig:wet_CVLCVR}, we show the constraints in the $(C_{V_L}^{(\ell)}, C_{V_R}^{(\ell)})$ plane, assuming real WCs and LFU in light leptons (see the discussion in the previous section). As anticipated from Eqs.~\eqref{eq:Blnu} and \eqref{eq:BPlnu}, the branching ratios of the fully leptonic $B\to \mu \nu$ and semileptonic $B\to \pi \ell \nu$ decay modes are sensitive to perpendicular directions in this plane, the former to the axial direction and the latter to the vectorial direction. The constraint from $B\to e \nu$ is not competitive in this plane, and we do not show it. On the other hand, the branching ratios of the semileptonic decay modes to final states with vector mesons are sensitive to both directions; see Eq.~\eqref{eq:BVlnu}. As already anticipated in Ref.~\cite{Bernlochner:2021rel} and further demonstrated in Ref.~\cite{Leljak:2023gna}, we observe a tension in the global fit at the level of $\sim2\sigma$ with respect to the SM point. The tension is a consequence of the measured differential spectra of $B\to \omega \ell \nu$ and $B\to \rho \ell \nu$ being consistently below the SM predictions (see e.g.~\cite{Bernlochner:2021rel}). A negative interference with the SM contribution is needed in order to explain the data, hence $C_{V_L}^{(\ell)}<0$ is preferred. Notice that two degenerate best-fit regions are found, the first one being close to the SM point, where NP is a small correction, while the second one represents the region in which NP is almost canceling the SM contribution. Note, however, that the best-fit regions are in slight tension with the current constraint from $B\to \mu \nu$. This constraint, however, comes with the caveat that the semileptonic $B\to \rho \ell \nu$ and $B\to \omega \ell \nu$ represent a part of the background events in the analysis of $B\to \mu \nu$. These, in turn, depend on the WCs considered here. To account for this properly, a WET analysis would have to be performed already at the level of the experimental analysis. At this stage, it is hard to quantify this effect properly. However, moving to the best-fit point would decrease these backgrounds and potentially slightly increase the significance of the signal. This would further worsen the disagreement between the region preferred by $B\to V \ell \nu$ and that preferred by $B\to \mu \nu$. 

The $C_{V_L}^{(\ell)}=- C_{V_R}^{(\ell)}$ direction is not sensitive to $B \to \pi \ell \nu$ decays and is preferred by the global fit in the $(C_{V_L}^{(\ell)}, C_{V_R}^{(\ell)})$ plane as a solution to a mild $B \to \{\rho,\omega\} \ell \nu$ discrepancy. In the upper right plot of Figure~\ref{fig:wet_CVLCVR}, we show the contours in the complex plane of this direction. Although a sizable imaginary part of the WCs is allowed by the global fit, a negative $\mathrm{Re} C_{V_L}^{(\ell)}=- \mathrm{Re} C_{V_R}^{(\ell)}$ is needed so as to interfere with the SM, as discussed above, destructively. For this reason, in the following, we will only assume NP that is aligned with the phase of the SM contributions.

Finally, in the bottom plot of Figure~\ref{fig:wet_CVLCVR} we show the dependence of $\Delta \chi^2 \equiv \chi^2 - \chi^2_{\mathrm{min.}}$ on the assumed direction of real $C_{V_L}^{(\ell)}=- C_{V_R}^{(\ell)}$. The constraint from $B\to V \ell \nu$ is again showcasing the $\sim2.5\sigma$ tension with respect to the SM, whereas the constraint from $B\to \mu \nu$ is in slight tension. Furthermore, we assume two well-motivated assumptions about the WCs for the $\tau$ lepton flavor -- either lepton flavor universality in $C_{V_R}^{(\tau)} = C_{V_R}^{(\ell)}$, or lepton flavor universality in both $C_{V_L}^{(\tau)} = C_{V_L}^{(\ell)}$ and $C_{V_R}^{(\tau)} = C_{V_R}^{(\ell)}$ (see Section~\ref{sec:SMEFT} for the reasoning behind these assumptions). The constraints from $B\to \tau \nu$ further challenge the region preferred by $B\to V \ell \nu$ and will play an important role in the scenarios considered in Section~\ref{sec:SMEFT}.

\begin{figure}[t]
     \centering
     \begin{subfigure}[b]{0.49\textwidth}
         \centering
         \includegraphics[width=\textwidth]{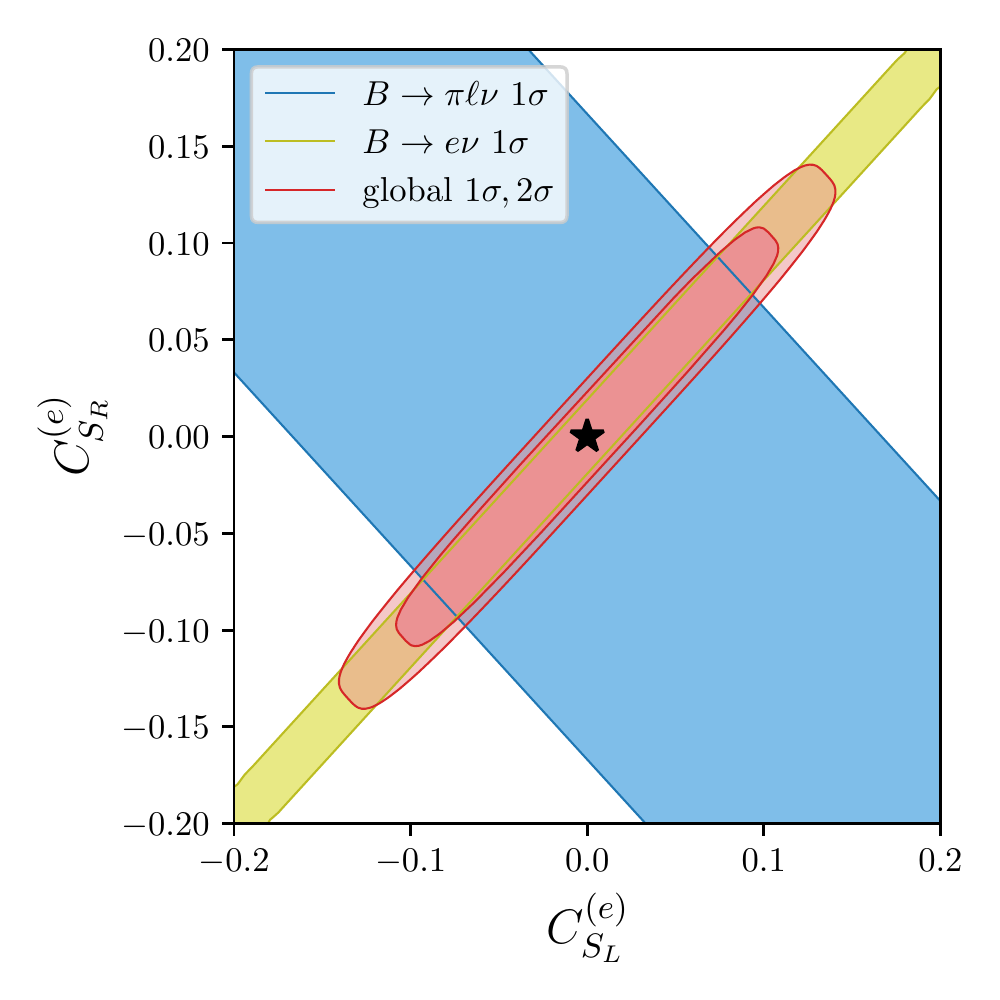}
     \end{subfigure}~
     \begin{subfigure}[b]{0.49\textwidth}
         \centering
         \includegraphics[width=\textwidth]{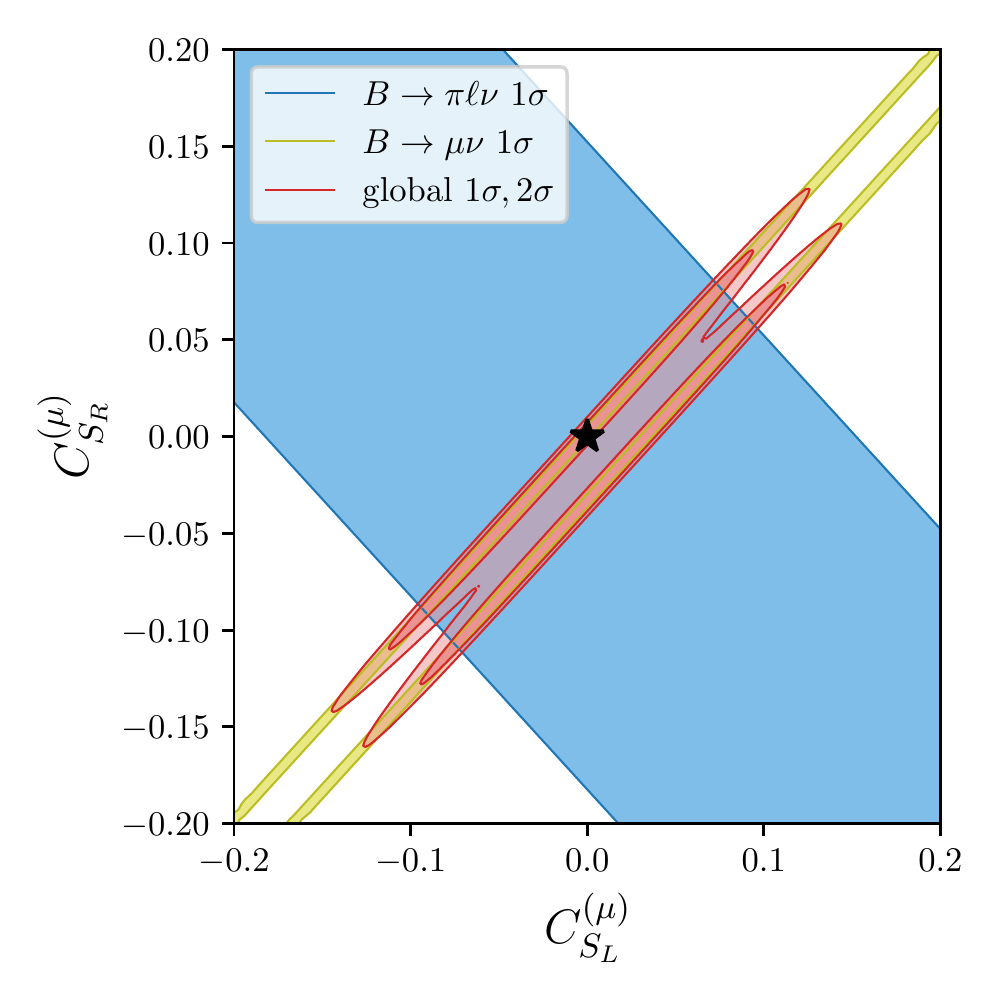}
     \end{subfigure}
        \caption{
        2D contours in the scenario $(C_{S_L}, C_{S_R})$, assuming NP either in electrons (left) or in muons (right).
} 
        \label{fig:wet_CSLCSR}
\end{figure}

We now turn our attention to the scalar operators defined in Eq.~\eqref{eq:OSLOSR}. In Figure~\ref{fig:wet_CSLCSR}, we show the constraints in the plane of $(C_{S_L}, C_{S_R})$, assuming NP in either electrons (left plot) or in muons (right plot).  As anticipated from Eqs.~\eqref{eq:Blnu} and \eqref{eq:BPlnu}, the fully leptonic decay modes are sensitive to the pseudoscalar direction, whereas the semileptonic decay modes to a pseudoscalar final state meson are sensitive to the perpendicular (scalar) direction. The constraints from $B\to V \ell \nu$ are not competitive in this plane, and hence we omit them in the plots.\footnote{Note that $B\to V \ell \nu$ decays also do not show tension in the plane of the scalar WCs, as the interference of these with the SM contribution is chirality suppressed.} Note that here we relax the assumption of lepton flavor universality, which in the case of $B\to \pi \ell \nu$ amounts to assuming NP only in one of the lepton channels, whereas still accounting for the SM prediction for the other channel. Due to the chirality enhanced sensitivity of $B\to \ell \nu$, the pseudoscalar direction is much better constrained compared to the scalar one. Furthermore, the branching ratio of $B\to \mu \nu$ is significantly better constrained than $B\to e \nu$, rendering the constraints in the muon scenario more stringent than the electron scenario. In the muon scenario, we again observe two minima, one of which amounts to a significant cancelation between (large) NP and the SM contributions to $B\to \mu \nu$. Finally, we comment that in this plane, there is less ambiguity expected in the contours from $B\to \ell \nu$ due to the dependence of the backgrounds on the WCs, i.e., $B\to \pi \ell \nu$ is blind to the direction to which $B\to \ell \nu$ is sensitive, whereas $B\to V \ell \nu$ are only marginally sensitive to the scalar WCs at the current experimental precision. 

Next, in Table~\ref{tab:WET-1D-bounds}, we collect the $2\sigma$ bounds on the WET WCs in the $b\to u \ell \nu$ sector, as organized by the Lorentz structure and parity of their respective operators. These results were obtained by considering one nonzero combination of WET WCs at a time. In all cases, except for the pseudoscalar coefficient, we assume LFU. In each case, a single process (we group the vector final states into $V=(\rho, \omega)$) dominates the global fit, as can be seen by comparing the $2\sigma$ regions from the processes, pointed out in each column of the table, with the global fit underneath. The constraint on the vector operator is completely dominated by measurements of the $B\to \pi \ell \nu$ branching ratios, as is already confirmed in Figure~\ref{fig:wet_CVLCVR}. The perpendicular, axial-vector direction is, on the other hand, most constrained by $B\to V \ell \nu$, where $V=(\omega, \rho)$, and shows a slight tension with the SM prediction at the level of $\sim2.5\sigma$. As can already be expected from Figure~\ref{fig:wet_CSLCSR}, the scalar Wilson coefficient is most constrained from measurements of $B\to \pi \ell \nu$, whereas the perpendicular, pseudoscalar direction is best constrained from fully leptonic modes $B\to \ell \nu$, with the constraint on the pseudoscalar coefficient being much tighter due to the chirally enhanced sensitivity of fully leptonic modes. Lastly, the current best bound on the Wilson coefficient of the tensor operator comes from $B\to V \ell \nu$, which can be understood from Eq.~\eqref{eq:BVlnu}. The tensor operator contributes constructively to $B\to V \ell \nu$, worsening the tension between predictions and experimental measurements. This effect is enough to render the $2\sigma$ constraint from $B\to V \ell \nu$ a factor of $\sim2$ better compared to the constraint from $B\to \pi \ell \nu$.

\begin{table}[]
\centering
\resizebox{\textwidth}{!}{%
\begin{tabular}{c|c|c|c|c|c}
$C_V^{\ell}$ & $C_A^{\ell}$ & $C_S^{\ell}$ & $C_P^{e}$ & $C_P^{\mu}$ & $C_T^{\ell}$ \\ \hline
\begin{tabular}[c]{@{}c@{}}$B\to \pi \ell \nu$:\\ $[-0.057, 0.108]$\end{tabular} & \begin{tabular}[c]{@{}c@{}}$B\to V \ell \nu$:\\ $[0.079, 0.397]$\end{tabular} & \begin{tabular}[c]{@{}c@{}}$B \to \pi \ell \nu$:\\ $[-0.178, 0.162]$\end{tabular} & \begin{tabular}[c]{@{}c@{}}$B\to e \nu$:\\ $[-0.027, 0.027]$\end{tabular} & \begin{tabular}[c]{@{}c@{}}$B\to \mu \nu$:\\ $[-0.008, 0.009]$\end{tabular} & \begin{tabular}[c]{@{}c@{}}$B\to V \ell \nu$:\\ $[-0.091, 0.095]$\end{tabular} \\ \hline
\begin{tabular}[c]{@{}c@{}}global:\\ $[-0.066, 0.094]$\end{tabular} & \begin{tabular}[c]{@{}c@{}}global:\\ $[0.039, 0.382]$\end{tabular} & \begin{tabular}[c]{@{}c@{}}global:\\ $[-0.178, 0.162]$\end{tabular} & \begin{tabular}[c]{@{}c@{}}global:\\ $[-0.027, 0.027]$\end{tabular} & \begin{tabular}[c]{@{}c@{}}global:\\ $[-0.008, 0.009]$\end{tabular} & \begin{tabular}[c]{@{}c@{}}global:\\ $[-0.09, 0.095]$\end{tabular}
\end{tabular}%
}
\caption{$2\sigma$ bounds on the vector, axial-vector, scalar, pseudoscalar, and tensor WCs in the WET. In each case, we point out the process that dominates the global fit (note that $V=(\rho, \omega)$). Each column represents a $1$D fit with only a single combination of WET WCs active at a time.} 
\label{tab:WET-1D-bounds}
\end{table}

\section{Global SMEFT analysis}
\label{sec:SMEFT}

In this section, we presume that the NP scale surpasses the EW scale, allowing the SMEFT to represent infrared (IR) physics adequately. By limiting the SMEFT expansion to the leading, dimension-6 order, we identify correlations among various low-energy observables. Moreover, starting with the minimal set of operators for $b \to u \ell \nu$ decays at the NP matching scale, significant effects are produced through the RG evolution down to the EW scale.

In Subsection~\ref{sec:SMEFT-setup}, we detail the relevant SMEFT operators and elucidate their matching to the WET operators. Subsequently, in Subsection~\ref{sec:compl}, we explore various observables that are correlated with $b \to u \ell \nu$ decays in the SMEFT context. Ultimately, in Subsection~\ref{sec:SMEFT-results}, we offer an extensive global study of these processes in the SMEFT.

\begin{table}[t]
    \centering
    \begin{tabular}{c|c}
        $Q_{lq}^{(3)}$ & $(\bar l_p\gamma_\mu\sigma_a l_r)(\bar q_s \gamma^\mu\sigma^a q_t)$ \\ \hline
        $Q_{ledq}$ & $(\bar l_p^j e_r)(\bar d_s q_{tj})$ \\
        $Q_{lequ}^{(1)}$ & $(\bar l_p^j e_r)\varepsilon_{jk}(\bar q_s^k u_t)$ \\
        $Q_{lequ}^{(3)}$ & $(\bar l_p^j \sigma_{\mu\nu} e_r)\varepsilon_{jk}(\bar q_s^k \sigma^{\mu\nu} u_t)$ \\ \hline
        $Q_{\phi q}^{(3)}$ & \rule{0pt}{2.9ex}$(\phi^\dagger i \protect\olr{D}_\mu^a \phi) (\bar{q}_p \sigma_a \gamma^\mu q_r)$ \\
        $Q_{\phi ud}$ & $(\tilde{\phi}^\dagger i D_\mu \phi) (\bar{u}_p \gamma^\mu d_r)$
        
    \end{tabular}~~~~
    \begin{tabular}{c|c}
        $Q_{lq}^{(1)}$ & $(\bar l_p\gamma_\mu l_r)(\bar q_s \gamma^\mu q_t)$ \\ \hline
        $Q_{\phi q}^{(1)}$ & \rule{0pt}{2.9ex}$(\phi^\dagger i \protect\olr{D}_\mu \phi) (\bar{q}_p \gamma^\mu q_r)$ \\

    \end{tabular}
    \caption{SMEFT operators at dimension 6 ($\Delta B =0$) contributing to (semi)leptonic $b\to u l \nu$ processes at the tree level (left) and closely related operators (right). Pauli matrices are denoted as $\sigma^a$, $\varepsilon$ is the totally anti-symmetric tensor in the $SU(2)_L$ space, $\protect\olr{D}_\mu \equiv D_\mu - \protect\olarrow{D}_\mu$ and $\protect\olr{D}_\mu^a \equiv \sigma^a D_\mu - \protect\olarrow{D}_\mu\sigma^a$.
    }
    \label{tab:SMEFToperators}
\end{table}

\subsection{Setup}
\label{sec:SMEFT-setup}

Above the EW scale, we use the following SMEFT effective Lagrangian at mass dimension~6 to parameterize model-independent effects of heavy NP,
\begin{equation}
\mathcal{L_{\mathrm{eff}}} = \mathcal{L_{\mathrm{SM}}} + \sum_{Q_i=Q_i^\dagger} \frac{C_i}{\Lambda^2} Q_i + \sum_{Q_i\neq Q_i^\dagger} \left( \frac{C_i}{\Lambda^2} Q_i + \frac{C_i^\ast}{\Lambda^2} Q_i^\dagger \right) \,,
\label{eq:SMEFT}
\end{equation}
where $Q_i$ are local effective operators in the Warsaw basis~\cite{Grzadkowski:2010es}, $C_i$ are WCs, and $\Lambda$ is the cutoff scale. In Table~\ref{tab:SMEFToperators}, we collect the subset of dimension-6 operators we focus on in this work: those that contribute at the tree level to $b\to u \ell \nu$ processes and the closely related operators.\footnote{That is, operators contributing to the same set of complementary observables and (or) are generated when integrating a perturbative UV model.} The operators $Q_{lq}^{(3)}, Q_{ledq}, Q_{lequ}^{(1)}$ and $Q_{lequ}^{(3)}$ contribute to $b\to u \ell \nu$ as contact interactions, whereas $Q_{\phi q}^{(3)}$ and $Q_{\phi ud}$ modify the left- and right-handed $W$ couplings with quarks, respectively. Already at this stage, we omit the operator $Q_{\phi l}^{(3)}$, which could contribute to $b\to u \ell \nu$ through a modified $W$ coupling to leptons, however, these operators are tightly constrained from other processes (see Section~\ref{sec:compl}).

For completeness, the tree level matching of the operators in Table~\ref{tab:SMEFToperators} to the WET Hamiltonian in Eq.~\eqref{eq:WET}, assuming the down-diagonal quark mass basis, is\footnote{Throughout this work, we consider a set of SMEFT operators that do not modify $v$ and $G_F$.}
\begin{align}
C_{V_L}^{(l)} &=
-\frac{V_{ud}}{V_{ub}} \frac{v^2}{\Lambda^2} \left[C_{lq}^{(3)}\right]_{ll13} +  \frac{V_{ud}}{V_{ub}} \frac{v^2}{\Lambda^2} \left[C_{\phi q}^{(3)}\right]_{13}\,,
&
C_{V_R}^{(l)} &=
\frac{1}{2 V_{ub}} \frac{v^2}{\Lambda^2} \left[C_{\phi u d}\right]_{13} \,,\label{eq:CVLCVR}
\\
C_{S_L}^{(l)} &=
-\frac{1}{2 V_{ub}} \frac{v^2}{\Lambda^2} \left[C_{lequ}^{(1)}\right]_{ll31}^\ast \,,
&
C_{S_R}^{(l)} &=
-\frac{V_{ud}}{2 V_{ub}} \frac{v^2}{\Lambda^2} \left[C_{ledq}\right]_{ll31}^\ast  \,,\label{eq:CSLCSR}
\\
C_{T}^{(l)} &=
-\frac{1}{2 V_{ub}} \frac{v^2}{\Lambda^2} \left[C_{lequ}^{(3)}\right]_{ll31}^\ast \,.\label{eq:CT}
\end{align}
Interestingly, at dimension-6 SMEFT, the left-handed vector operator can be generated either through a contact interaction $C_{lq}^{(3)}$ or through a modification of the $W$ coupling with left-handed quarks via $C_{\phi q}^{(3)}$, whereas the right-handed vector operator can only be generated through a modification of the $W$ coupling with right-handed quarks $C_{\phi u d}$.\footnote{The non-universal right-handed vector operator in the WET is generated by the dimension-8 SMEFT operator $(\tilde \phi^\dagger \sigma^a \phi)(\bar u \gamma^\mu d)(\bar l \gamma_\mu \sigma^a l)$.}

Note that, assuming the standard CKM parameterization, $V_{ub}$ in Eqs.~\eqref{eq:CVLCVR} - \eqref{eq:CT} is complex and carries an imaginary part which is almost $3$ times as large as its real part. Assuming real SMEFT WCs would thus result in WET WCs with substantial imaginary parts, firmly selecting a direction, for e.g., the upper right plot on Figure~\ref{fig:wet_CVLCVR}. In order to circumvent the computational complexity of assuming complex WCs in the SMEFT, we choose a direction in the complex plane of each WC that is aligned with the phase of $V_{ub}$ in the following way:
\begin{equation}
\label{eq:SMEFTredefine}
    C_i = \frac{V_{ub}}{|V_{ub}|} \tilde{C}_i \,,\qquad \tilde{C}_i \in \mathbb{R} \,.
\end{equation}
It is straightforward to see that in Eqs.~\eqref{eq:CVLCVR} - \eqref{eq:CT} the $V_{ub}$ factor will cancel out. However, we keep its absolute value so as not to change the magnitude of WCs. In the following, we will present results in terms of $\tilde{C}_i$, unless otherwise stated.

As for the numerical analysis, both in this and in the following section we use the same setup as already described in Section \ref{sec:WET}. Most of the complementary constraints discussed in the following are already a part of the public \texttt{smelli}~\cite{Aebischer:2018iyb,Stangl:2020lbh} likelihood, with the exception of high-mass Drell-Yan tails, which we add based on their implementation in \texttt{flavio}~\cite{Greljo:2022jac}. As already mentioned in Section \ref{sec:WET}, we take into account the dependence of the theory uncertainties on the NP parameters~\cite{Altmannshofer:2021qrr,Greljo:2022jac}. We choose the initial scale at which the SMEFT operators are defined to be $\Lambda=1~\mathrm{TeV}$, and rely on \texttt{wilson}~\cite{Aebischer:2018bkb} for the running and matching\footnote{Note, that the current official version of \texttt{wilson} is neglecting two insertions of vertex corrections in the matching of the SMEFT to the WET, which is necessary for some complementary constraints, see Section~\ref{sec:complementary_Bmixing}. We have added these neglected terms in a private version of \texttt{wilson}, and they will be available in a future official release.} of the WCs, both above and below the EW scale, and on the Wilson coefficient exchange format (\texttt{WCxf})~\cite{Aebischer:2017ugx} to represent the Wilson coefficients and to fix the EFT bases.

\subsection{Complementary constraints}
\label{sec:compl}

In this section, we discuss the phenomena correlated to $b \to u \ell \nu$ transitions within the SMEFT framework. These include: neutral-current rare $b$ decays, high-mass Drell-Yan production $p p \to \ell^+ \ell^-$ and $p p \to \ell \nu$, $B^0 - \bar B^0$ oscillations, and EW gauge boson vertex corrections.

\subsubsection{Rare $b$ decays}
\label{sec:complementary_NC_Bdecays}

Due to $SU(2)_L$ relations, the operators $Q_{lq}^{(3)}, Q_{ledq}, Q_{\phi q}^{(3)}$, together with the related operators $Q_{lq}^{(1)}$ and $Q_{\phi q}^{(1)}$, enter the leptonic and semileptonic rare $B$ decays with the underlying transition $b\to d \ell \ell$ with no additional suppression, either as contact interactions or as modifications of the $Z$ boson couplings with quarks. This leads to important constraints from measurements of the leptonic branching ratios $B\to e e$, reported by LHCb~\cite{LHCb:2020pcv}, and $B\to \mu \mu$, reported by LHCb~\cite{LHCb:2021awg, LHCb:2021vsc}, CMS \cite{CMS:2022mgd} and ATLAS~\cite{ATLAS:2018cur}, and semileptonic branching ratios $B\to \pi e e$, the upper limit of which was reported by Belle~\cite{Belle:2008tjs}, $B\to \pi \mu \mu$, measured differentially by LHCb~\cite{LHCb:2015hsa}, and $B_s\to K^{\ast 0} \mu \mu$ observed by LHCb~\cite{LHCb:2018rym} with a significance of $3.4\sigma$. See Refs.~\cite{Bause:2022rrs, Greljo:2022jac} for dedicated studies of this sector in the WET. The related FCNC process $b\to d \nu \nu$ is sensitive to the operators $Q_{lq}^{(1)}, Q_{lq}^{(3)}$, $Q_{\phi q}^{(1)}$ and $Q_{\phi q}^{(3)}$. The upper limits on $B\to \pi \nu \nu$ and $B\to \rho \nu \nu$ have been determined by Belle using either hadronic~\cite{Belle:2013tnz} or semileptonic tagging~\cite{Belle:2017oht}.

In both $b\to d \ell \ell$ and $b\to d \nu \nu$ we expect unconstrained directions in the $(C_{lq}^{(1)}, C_{lq}^{(3)})$ and $(C_{\phi q}^{(1)}, C_{\phi q}^{(3)})$ pairs of WCs. Namely, in the case of contact interactions, there will be no contribution to $b\to d \ell \ell$ when aligned with the direction of $C_{lq}^{(1)} = -C_{lq}^{(3)}$, and no contribution to $b\to d \nu \nu$ in the perpendicular direction of $C_{lq}^{(1)} = C_{lq}^{(3)}$. These directions are also stable under RG effects. On the other hand, in the case of modified $Z$ vertices, both $b\to d \ell \ell$ and $b\to d \nu \nu$ have the same flat direction of $C_{\phi q}^{(1)} = -C_{\phi q}^{(3)}$ at the tree level, as this combination does not contribute to down-quark FCNCs. This relation is, however, mildly broken due to the different renormalization of the singlet and triplet operators. The effect is dominated by the $y_t$-enhanced contribution from the diagrams in which one of the Higgs legs is attached to a quark leg to form a loop, and a Higgs is emitted from the quark in the loop. Writing only the dominant parts of the RG equations (RGE) contributing to the effect relevant to our discussion, we have~\cite{Jenkins:2013wua}
\begin{equation}
    \label{eq:RGE:phiq1phiq3}
\begin{split}
    \left[\dot{C}_{\phi q}^{(1)}\right]_{pr} \propto \frac{3}{2}  \left[C_{\phi q}^{(1)}\right]_{pt} [Y_u^\dagger Y_u]_{tr} - \frac{9}{2} \left[C_{\phi q}^{(3)}\right]_{pt} [Y_u^\dagger Y_u]_{tr} \,, \\
    \left[\dot{C}_{\phi q}^{(3)}\right]_{pr} \propto \frac{1}{2}  \left[C_{\phi q}^{(3)}\right]_{pt} [Y_u^\dagger Y_u]_{tr} - \frac{3}{2} \left[C_{\phi q}^{(1)}\right]_{pt} [Y_u^\dagger Y_u]_{tr} \,,
\end{split}
\end{equation}
where $\dot{C}\equiv 16 \pi^2 \mu \frac{d}{d\mu}C$. This will be relevant for understanding Figure~\ref{fig:smeft_phiq1_phiq3_phiud}. 

Finally we comment on the $Q_{\phi u d}$ operator, which in principle runs into $Q_{\phi u}$ and $Q_{\phi d}$~\cite{Jenkins:2013wua}. These operators contribute to modified $Z$ boson couplings with right-handed quarks. However, these terms in the SMEFT RGE are $Y_d$ suppressed, rendering rare $b$ decays an inefficient probe of $Q_{\phi u d}$, when compared to bounds obtained from $b\to u \ell \nu$.

\subsubsection{High-mass Drell-Yan}

The constraints from measurements of Drell-Yan (DY) processes at high $p_T$ have been shown to be highly complementary to various low energy flavor processes, in particular in the sectors of rare $b$ decays $b\to s \ell \ell$ and $b\to d \ell \ell$~\cite{Greljo:2017vvb, Greljo:2022jac, Afik:2019htr}, charged-current $b \to c \tau \nu$ decays~\cite{Faroughy:2016osc, Greljo:2018tzh, Marzocca:2020ueu, Allwicher:2022gkm}, lepton flavor violating transitions~\cite{Angelescu:2020uug, Descotes-Genon:2023pen}, and in the charm sector~\cite{Fuentes-Martin:2020lea, Fajfer:2023nmz}. In this work, we study the complementarity between low-energy $b\to u \ell \nu$ processes and high-mass DY. 

High-mass DY processes are especially sensitive to contact interactions due to the favorable energy enhancement of the EFT amplitudes in the tails of high-$p_T$ distributions,\footnote{The $W$ and $Z$ vertex corrections do not exhibit the same energy enhancement and are negligible in the high-mass DY tails.} which can overcome the PDF suppression due to potential sea quarks in the initial state. An important property of the dependence of high-$p_T$ spectra to contact interactions is the lack of interference terms between different SMEFT operators. This means that the majority of the operators, barring the ones that interfere with the SM, will contribute to the spectra only constructively, and the extracted bounds will be free from unconstrained directions.

It is important to note here, that the high-mass DY tails are effectively probing energies up to the TeV scale. The EFT validity arguments suggest that the bounds derived from the high-mass tails are applicable primarily to models with energy scales beyond the TeV scale. Consequently, the constraints imposed by the DY tails are sensitive to a smaller subset of models when compared to the constraints obtained from low-energy observables. See e.g.~\cite{Greljo:2022jac} for a recent discussion.

We consider the latest data on the differential spectra of both neutral ($p p \to \ell \ell$) and charged current ($p p \to \ell \nu$) DY processes with light leptons in final states, both from CMS~\cite{CMS:2021ctt, CMS:2022yjm} and ATLAS~\cite{ATLAS:2020yat, ATLAS:2019lsy}. We have recently implemented both the theoretical predictions, including SMEFT at mass-dimension $6$, and the latest data into \texttt{flavio}, see Ref.~\cite{Greljo:2022jac} for details.

\subsubsection{$B^0$ meson mixing}
\label{sec:complementary_Bmixing}

The $C_{\phi q}^{(1)}$ and $C_{\phi q}^{(3)}$ WCs, generating flavor-changing modifications of the $Z$ boson couplings with left-handed quarks, contribute to $\Delta F=2$ processes at the tree-level, through two insertions of the modified vertex. However, it turns out that there is a numerically important contribution generated in the process of RG evolving these WCs from a high scale to a low scale. Namely, keeping only the terms relevant for this discussion, we have the following $y_t$ enhanced terms in the SMEFT RGE of the four-quark operators~\cite{Jenkins:2013wua}\footnote{Note, that the argument holds irrespective of the quark mass alignment. In the down-diagonal basis, Eq.~\eqref{eq:RGE:qq1qq3} directly produces the off-diagonal elements of the flavor tensors, as required by $\Delta F=2$ processes. In the up-diagonal basis, the necessary off-diagonal entries are instead generated in the process of matching SMEFT to the LEFT.}
\begin{equation}
    \label{eq:RGE:qq1qq3}
\begin{split}
    \left[\dot{C}_{q q}^{(1)}\right]_{prst} \propto 
    \frac{1}{2} [Y_u^\dagger Y_u]_{pr}  \left[C_{\phi q}^{(1)}\right]_{st} +
    \frac{1}{2} [Y_u^\dagger Y_u]_{st}  \left[C_{\phi q}^{(1)}\right]_{pr} \,, \\
    \left[\dot{C}_{q q}^{(3)}\right]_{prst} \propto 
    -\frac{1}{2} [Y_u^\dagger Y_u]_{pr}  \left[C_{\phi q}^{(3)}\right]_{st} 
    -\frac{1}{2} [Y_u^\dagger Y_u]_{st}  \left[C_{\phi q}^{(3)}\right]_{pr} \,.
\end{split}
\end{equation}
Note, that there is an important difference between the tree-level contributions due to modified $Z$ vertices, and the RGE-induced contributions to four-quark operators. Namely, in the case when $C_{\phi q}^{(1)} = - C_{\phi q}^{(3)}$, 
the tree-level SMEFT to WET matching does not induce modified $Z$-boson vertices. On the contrary, the RGE-induced contributions in Eq.~\eqref{eq:RGE:qq1qq3} contribute even for this combination of WCs, as can be seen in the sign difference between the singlet and triplet operator contributions in Eq.~\eqref{eq:RGE:qq1qq3}. As we will see in Section~\ref{sec:SMEFT-results}, the interplay of both effects is responsible for closing the contours from $\Delta F=2$ processes.

In this work we are interested in the $Q_{\phi q}^{(1)}$ and $Q_{\phi q}^{(3)}$ operators with the first and third generation quarks. Thus we expect important constraints from the $B^0 -\bar{B}^0$ mixing observables, namely the mass difference $\Delta M_d$ and the CP asymmetry $S_{\psi K}$ from the interference between $B^0 -\bar{B}^0$ mixing and the decay $B^0 \to \psi K_S$. We consider the experimental measurements of these quantities, as determined by the latest HFLAV average~\cite{HeavyFlavorAveragingGroup:2022wzx}.

\subsubsection{$W$ and $Z$ vertex corrections}

The modified couplings of the $W$ and $Z$ bosons to fermions undergo constraints from on-shell vector boson production and decay processes at both the LEP and LHC colliders. The leptonic vertex corrections, arising from the operator $Q_{\phi l}^{(3)}$, have been found to be bounded at a (sub)percent level, as discussed in~\cite{Efrati:2015eaa}. Consequently, within the current precision, the contribution of this operator to the variation of $b \to u \ell \nu$ transitions is deemed negligible.

The bounds on quark vertex corrections derived from on-shell processes, specifically the observables associated with $Z$ and $W$ boson pole measurements, exhibit relatively weak constraints compared to the complementary constraints originating from low-energy observables investigated in this study (see, for instance, Eq. (4.10) in~\cite{Efrati:2015eaa}). Furthermore, the evaluation of CKM-suppressed top quark decays at the LHC presents considerable challenges, as elaborated in~\cite{Faroughy:2022dyq}. Moreover, top quark flavor-changing neutral current decays ($t \to Z q$) and the corresponding $p p \to t + Z$ production, both sensitive to the $Q_{\phi q}^{(3)}$ operator, impose additional constraints that are also subdominant in comparison to those derived from low-energy observables discussed above~\cite{ATLAS:2023qzr, Durieux:2014xla, Cremer:2023gne}. For example, the ATLAS limit on $\mathcal{B}(t \to Z u) < 6.2\times 10^{-5}$ at $95\%$\,CL~\cite{ATLAS:2023qzr} translates as $|\tilde C^{(3)}_{\phi q}| \lesssim 0.2$, which is two orders of magnitude worse than the scale in Figure~\ref{fig:smeft_phiq1_phiq3_phiud}.

\subsection{Results}
\label{sec:SMEFT-results}

In this subsection we present the results of a SMEFT analysis, focusing on $b\to u\ell\nu$ processes and the important complementary constraints discussed in the previous subsection. All of the results presented here assume a minimalistic flavor structure of the SMEFT operators, aligned with a maximal expected effect on $b\to u\ell\nu$.  From the results we have presented and discussed in the WET (see Section~\ref{sec:WET}), we pinpoint groups of SMEFT operators in which we expect interesting correlations to appear. 

Firstly, we consider the SMEFT operators, which match onto the vector WET operators, since we have seen interesting correlations among them already in the WET (see Figure~\ref{fig:wet_CVLCVR}). For the right-handed vector operator $O_{V_R}$ there is only a single SMEFT operator generating it at dimension $6$, the $Q_{\phi u d}$. As for the left-handed vector operator $O_{V_L}$, we consider either the contact interaction operator $Q_{lq}^{(3)}$, together with its related operator $Q_{lq}^{(1)}$, or the vertex correction operator $Q_{\phi q}^{(3)}$, together with its related operator $Q_{\phi q}^{(1)}$. This leaves us with two groups of $3$ SMEFT operators to explore, and to clearly present the results we resort to either profiling over one of the directions, or projecting onto a well-motivated plane in the $3$-dimensional space. 

\begin{figure}[t]
     \centering
     \begin{subfigure}[b]{0.49\textwidth}
         \centering
         \includegraphics[width=\textwidth]{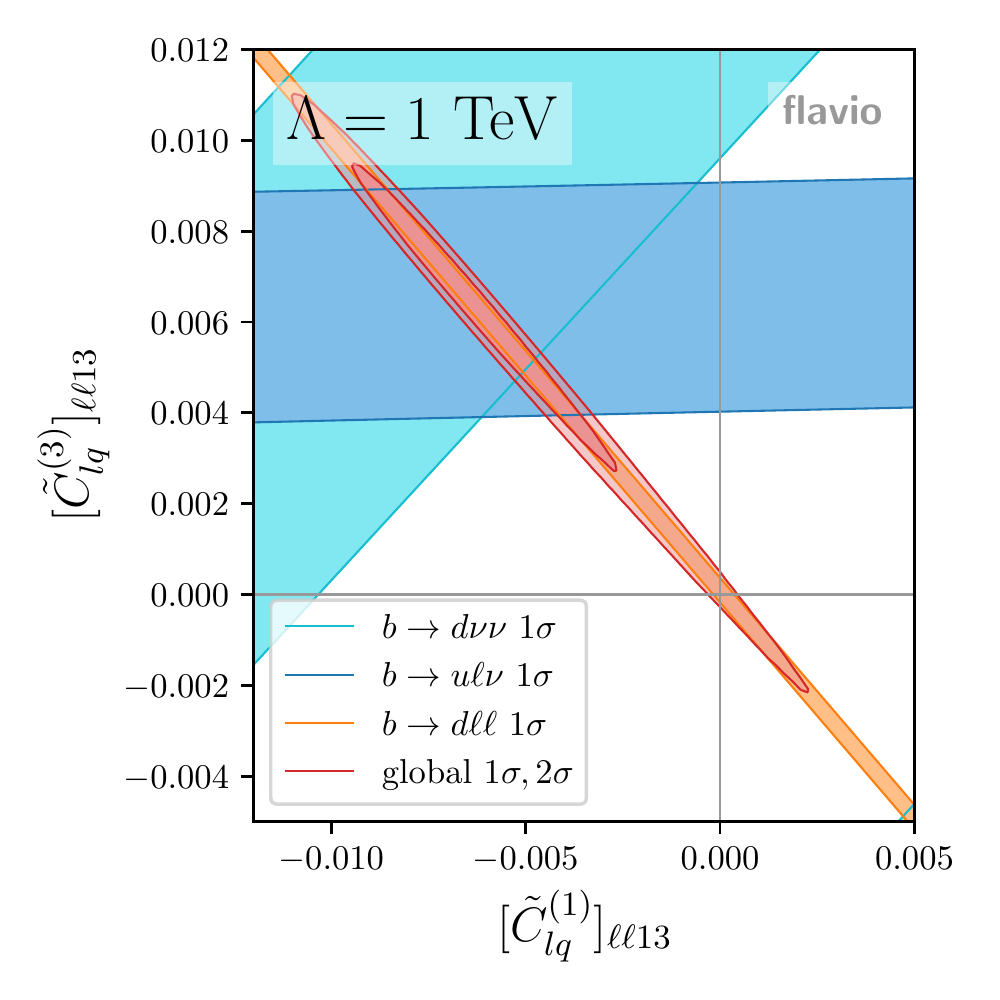}
     \end{subfigure}~
     \begin{subfigure}[b]{0.49\textwidth}
         \centering
         \includegraphics[width=\textwidth]{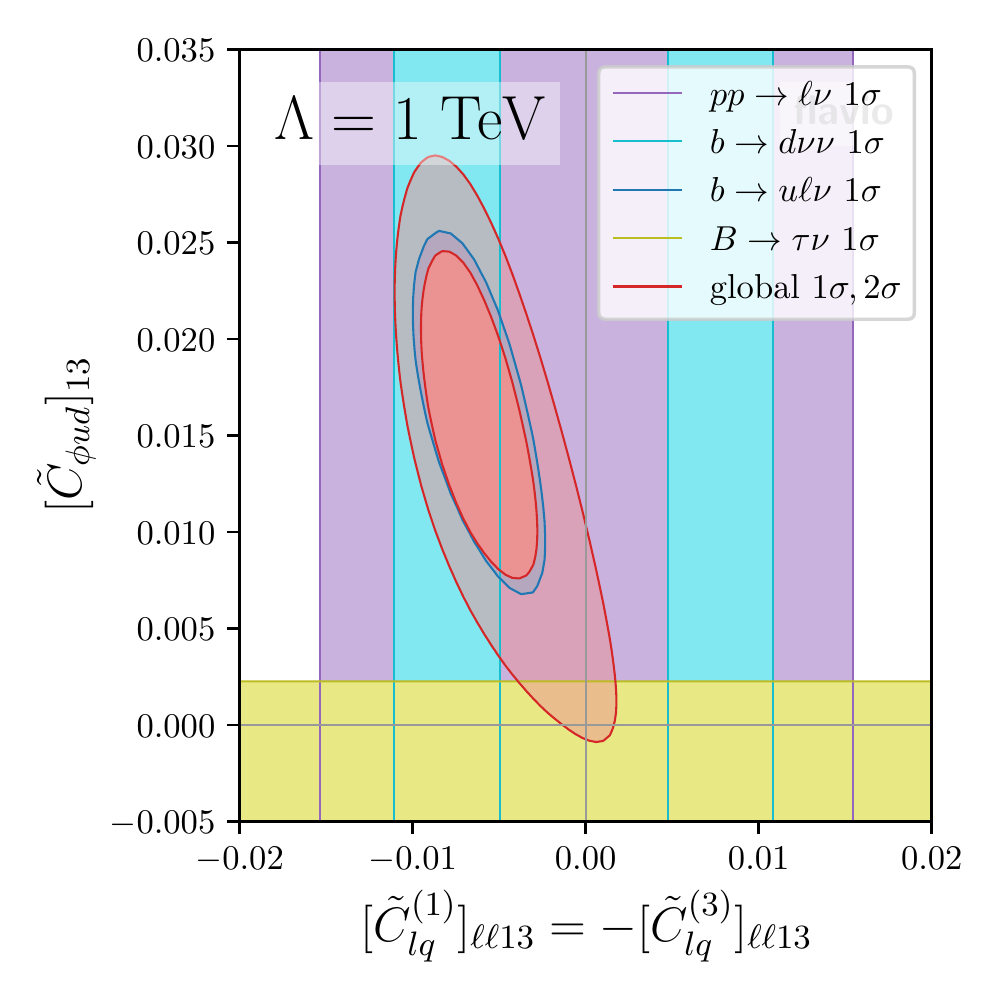}
     \end{subfigure}
        \caption{
        \textbf{Left:} 2D contours in the scenario $([\tilde{C}_{lq}^{(1)}]_{\ell\ell 13}, [\tilde{C}_{lq}^{(3)}]_{\ell\ell 13})$, profiled over $[\tilde{C}_{\phi ud}]_{13}$. 
        \textbf{Right:} 2D contours in the scenario $([\tilde{C}_{lq}^{(1)}]_{\ell\ell 13}=-[\tilde{C}_{lq}^{(3)}]_{\ell\ell 13}, [\tilde{C}_{\phi ud}]_{13})$. See Section~\ref{sec:SMEFT-results} for details.
        }
        \label{fig:smeft_lq1_lq3_phiud}
\end{figure}

{\bf Group I} --- In Figure~\ref{fig:smeft_lq1_lq3_phiud} we focus on the group of WCs $(\tilde C_{lq}^{(1)},\tilde C_{lq}^{(3)}, \tilde C_{\phi ud})$. The left plot shows the constraints in the $([\tilde{C}_{lq}^{(1)}]_{\ell\ell13}, [\tilde{C}_{lq}^{(3)}]_{\ell\ell13})$ plane, profiling over the $[\tilde C_{\phi ud}]_{13}$ direction. We emphasize, that the $b\to u l \nu$ processes are the only ones considered here sensitive to this direction. In fact, allowing for the combination of $C_{V_L}$ and $C_{V_R}$ to be generated in this scenario, the exclusive semi-leptonic $B$ decays are in tension with the SM at the level of $\sim 2.5\sigma$, as already anticipated in Figure~\ref{fig:wet_CVLCVR}. This seems to be further supported by the $b\to d\nu\nu$ constraint, exhibiting a slight tension with the SM at the level of $\sim 1.5\sigma$. By far the most constraining processes in this plane are the FCNC $b\to d \ell \ell$ processes, which are however insensitive in the direction of $\tilde C_{lq}^{(1)} = - \tilde C_{lq}^{(3)}$ unconstrained from these processes, resulting in a narrow, elongated global fit. The global fit exhibits a slight tension with the SM at the level of $\sim1.5\sigma$. The reason that the tension is weaker than with exclusive $b\to u \ell \nu$ alone is that the global fit also includes the leptonic decays $B\to l \nu$. As already anticipated in Figure~\ref{fig:wet_CVLCVR}, $[\tilde C_{\phi ud}]_{13}$ generates a universal right-handed vector operator, leading to important constraints not only from $B\to \ell \nu$ but also from $B\to \tau \nu$. Notably, the bounds obtained from Drell-Yan tails are not competitive in this scenario and, thus are not represented in the plot.

The right plot in Figure \ref{fig:smeft_lq1_lq3_phiud} depicts the bounds obtained for the case aligned with the $b\to d\ell\ell$ flat direction, where $\tilde C_{lq}^{(1)} = -\tilde C_{lq}^{(3)}$, against $\tilde C_{\phi ud}$. In this scenario, the stringent bounds from FCNC $b\to d \ell \ell$ processes are absent, allowing more freedom for the global fit. The contours from $b\to u \ell \nu$ processes, including both semileptonic and leptonic transitions with light leptons, again show tension with the SM point. This is however challenged by the $B\to \tau \nu$ constraint. We also show the complementary constraints from $b\to d \nu \nu$ processes and charged current high-mass DY tails, both of which only mildly impact the global fit. However, the global fit, mildly incompatible with the SM, will be further challenged by future measurements of these complementary processes.

\begin{figure}[t]
     \centering
     \begin{subfigure}[b]{0.49\textwidth}
         \centering
         \includegraphics[width=\textwidth]{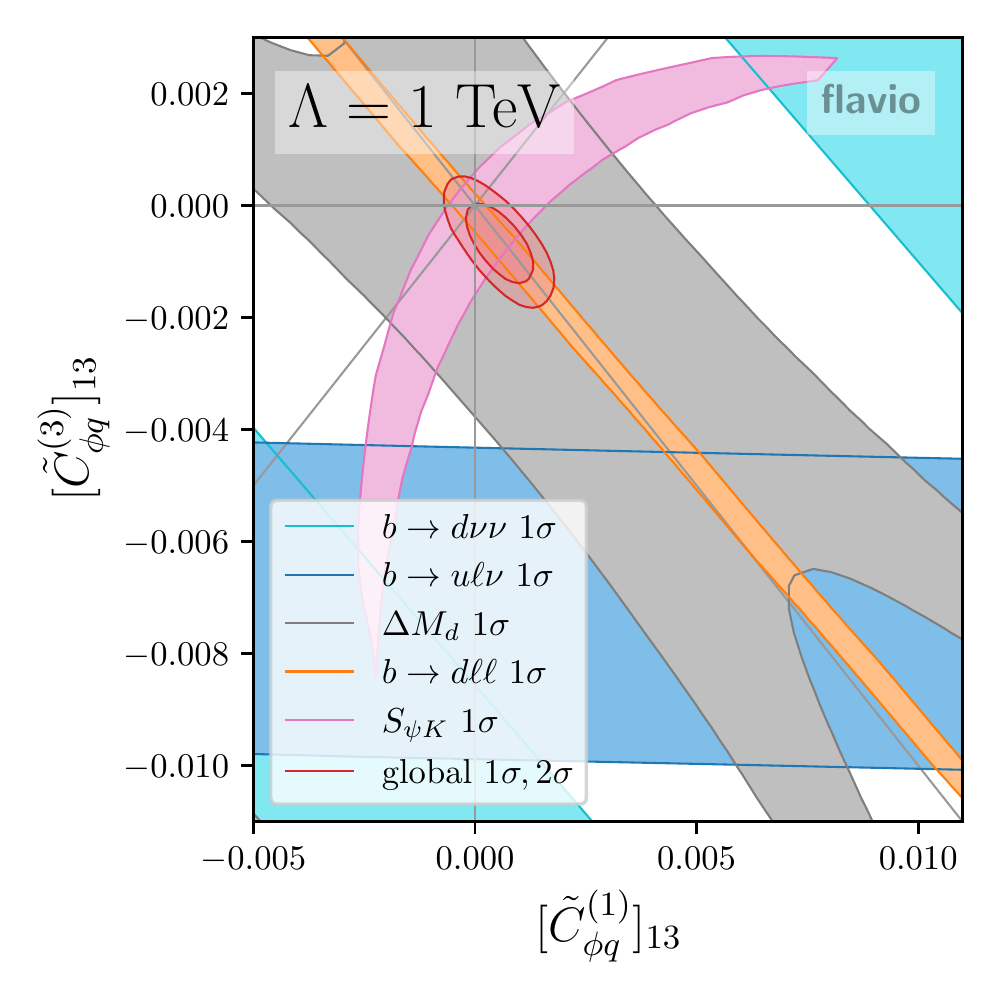}
     \end{subfigure}~
     \begin{subfigure}[b]{0.49\textwidth}
         \centering
         \includegraphics[width=\textwidth]{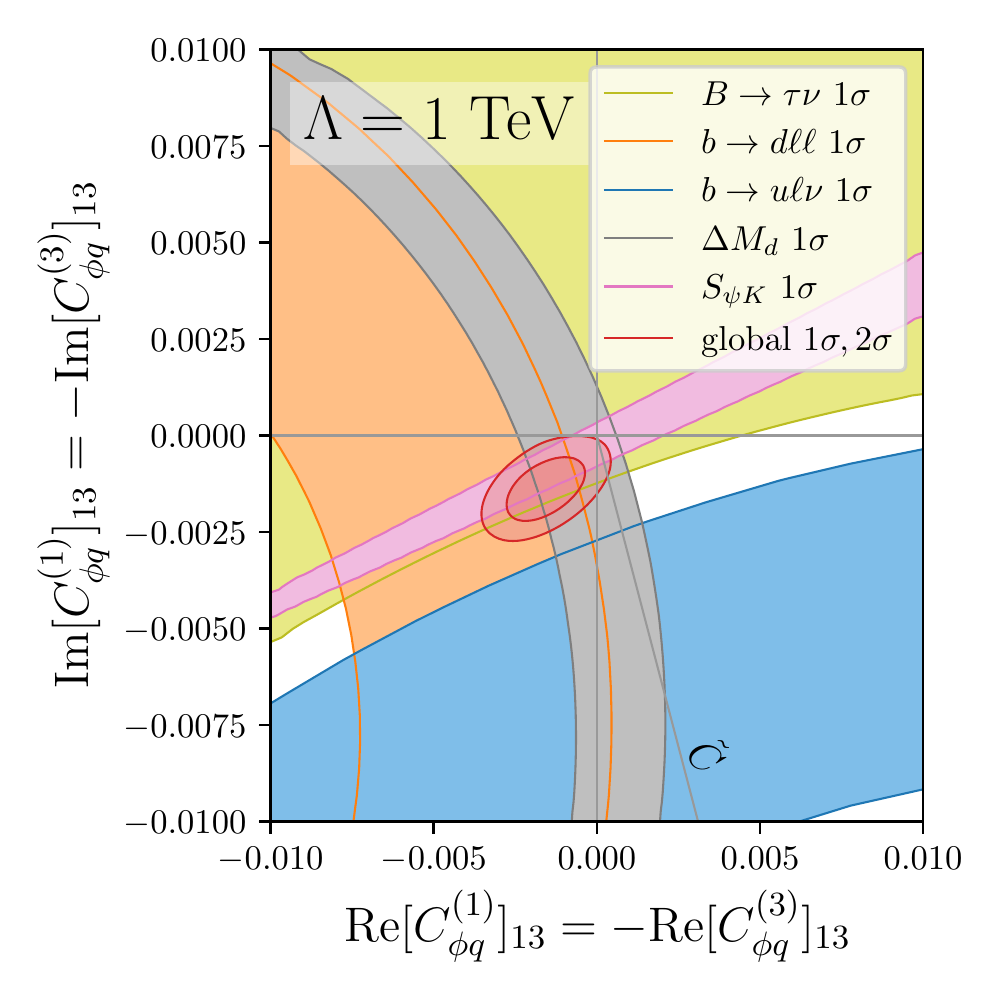}
     \end{subfigure}
        \caption{
        \textbf{Left:} 2D contours in the scenario $([\tilde{C}_{\phi q}^{(1)}]_{13}, [\tilde{C}_{\phi q}^{(3)}]_{ 13})$, profiled over $[\tilde{C}_{\phi ud}]_{13}$. 
        \textbf{Right:}  Contours assuming the direction of $C_{\phi q}^{(1)} = -C_{\phi q}^{(3)}$ and allowing for arbitrary imaginary parts, still profiling over $[\tilde{C}_{\phi ud}]_{13}$. See surrounding text for details.
        }
        \label{fig:smeft_phiq1_phiq3_phiud}
\end{figure}

{\bf Group II} --- In Figure~\ref{fig:smeft_phiq1_phiq3_phiud} we turn our attention to the group of WCs $(\tilde C_{\phi q}^{(1)},\tilde C_{\phi q}^{(3)}, \tilde C_{\phi ud})$, constituting only left- and right-handed quark vertex corrections. In the left panel of the figure, the constraints on the $(\tilde C_{\phi q}^{(1)},\tilde C_{\phi q}^{(3)})$ parameter space are presented, profiling over $\tilde C_{\phi ud}$. The $b\to u \ell \nu$ contours exhibit a tension with the SM, for the same reason as already discussed in the previous case. However, the situation changes drastically in relation to the complementary constraints. As discussed in Subsection~\ref{sec:complementary_NC_Bdecays}, both $b\to d \ell \ell$ and $b\to d \nu \nu$ now constrain the same direction, with the former being vastly more constraining, rendering the latter irrelevant for the global fit. Notice also the misalignment between the $b\to d \ell \ell$ contour and the direction of $\tilde{C}_{\phi q}^{(1)} = -\tilde{C}_{\phi q}^{(3)}$, as anticipated from the RG effects discussed in Subsection~\ref{sec:complementary_NC_Bdecays}. Furthermore, as anticipated in Subsection~\ref{sec:complementary_Bmixing}, neutral $B$-meson mixing constraints are important in this plane from two effects: two insertions of the modified $Z$-vertices at the tree level, and RGE-induced contributions to the four quark operators, as discussed in Subsection \ref{sec:complementary_Bmixing}. The $\Delta M_d$ constraint is more relaxed, however the mixing-induced CP asymmetry $S_{\psi K}$ is highly constraining, dominating and closing the global fit together with $b\to d \ell \ell$. This is due to our choice of basis in Eq.~\eqref{eq:SMEFTredefine}, where a substantial imaginary part is imposed on the coefficients $C$, which is, contrary to the $b\to u \ell \nu$ sector, not canceled in the matching to other sectors. 

This assumption is relaxed in the upper right plot of Figure~\ref{fig:smeft_phiq1_phiq3_phiud}, which momentarily omits the use of the $\tilde{C}$ basis (Eq.~\eqref{eq:SMEFTredefine}). We choose the direction of $C_{\phi q}^{(1)} = -C_{\phi q}^{(3)}$, and present contours in the complex plane of this scenario. We still profile over only the $\tilde{C}_{\phi u d} \in \mathbb{R}$ direction, to reduce the computational complexity, however, this has no meaningful impact on the results. The $b\to u \ell \nu$ contours prefer an imaginary part of the considered WCs, further supporting our choice of basis in Eq.~\eqref{eq:SMEFTredefine}, especially when considering the direction of $\tilde{C}$ as overlaid on the plot. The complementary constraints from $b\to d \ell \ell$ and $\Delta M_d$ could in principle support the tension exhibited by the $b\to u \ell \nu$ processes, however, note that there is no region in which the latter would be compatible with the stringent constraint from the mixing-induced CP asymmetry $S_{\psi K}$.

\begin{figure}[t]
     \centering
     \begin{subfigure}[b]{0.49\textwidth}
         \centering
         \includegraphics[width=\textwidth]{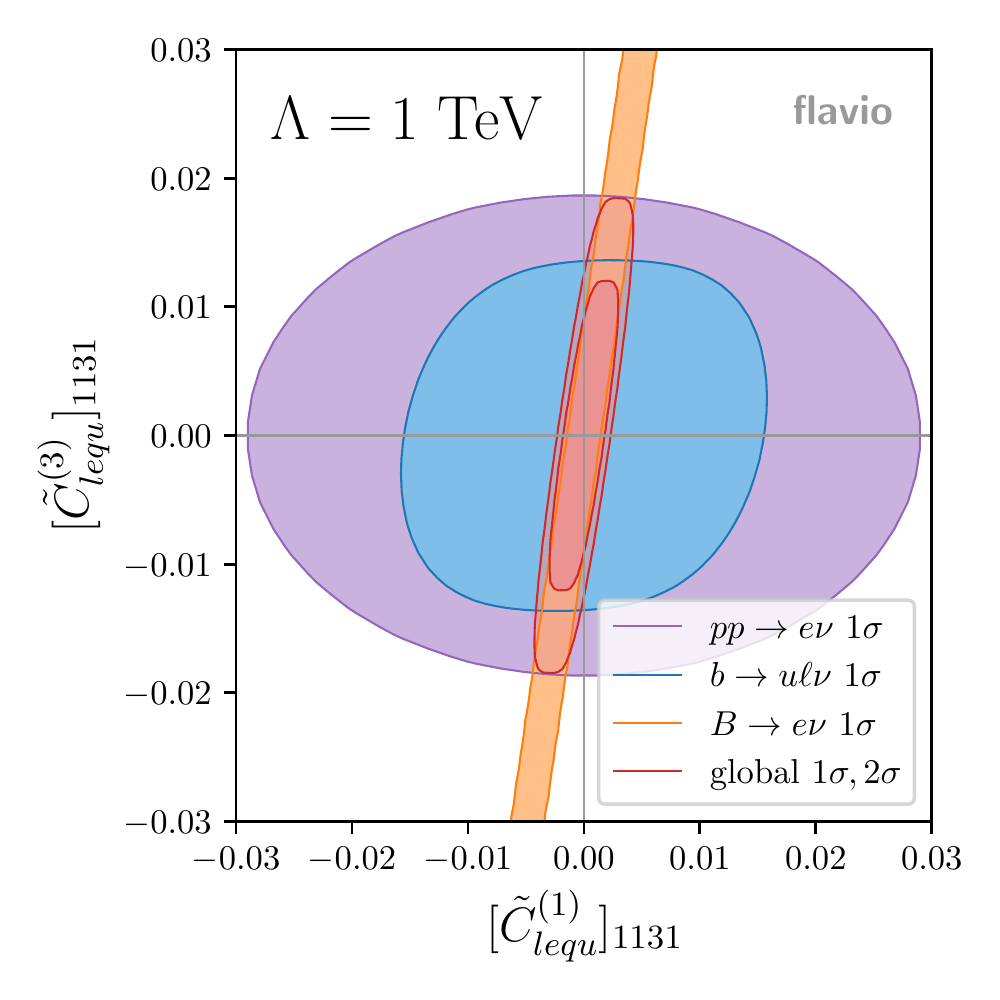}
     \end{subfigure}~
     \begin{subfigure}[b]{0.49\textwidth}
         \centering
         \includegraphics[width=\textwidth]{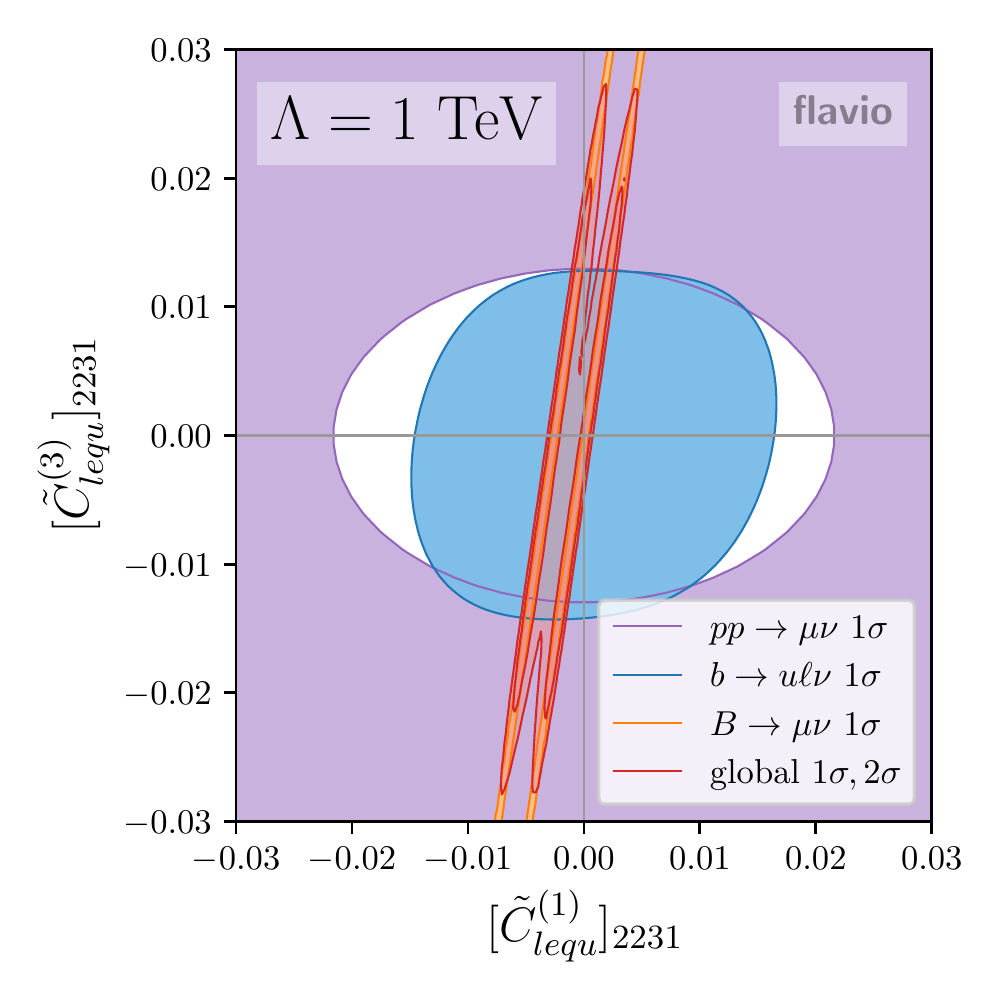}
     \end{subfigure}
        \caption{
        \textbf{Left:} 2D contours in the scenario $([\tilde{C}_{lequ}^{(1)}]_{1131}, [\tilde{C}_{lequ}^{(3)}]_{1131})$ (NP only in electrons).
        \textbf{Right:} 2D contours in the scenario $([\tilde{C}_{lequ}^{(1)}]_{2231}, [\tilde{C}_{lequ}^{(3)}]_{2231})$ (NP only in muons). See surrounding text for details.}
        \label{fig:smeft_lequ1_lequ3}
\end{figure}

{\bf Other operators} --- Lastly, we consider the scalar and tensor operators $Q_{lequ}^{(1)}$, $Q_{lequ}^{(3)}$ and $Q_{ledq}$. Motivated by the discussion in the following Section~\ref{sec:models}, we group them so that we first consider the pair $(Q_{lequ}^{(1)}$, $Q_{lequ}^{(3)})$, and then consider $Q_{ledq}$ separately. As we expect the bounds on these to be dominated by fully leptonic $B$ decays, we also separate each scenario by the lepton flavor (see Section~\ref{sec:WET}), omitting our assumption of LFU. 

In Figure~\ref{fig:smeft_lequ1_lequ3} we study the operators $Q_{lequ}^{(1)}$, $Q_{lequ}^{(3)}$ contributing to $b\to u \ell \nu$, assuming NP only in electrons (left plot) or only in muons (right plot). The $Q_{lequ}^{(1)}$ and $Q_{lequ}^{(3)}$ operators match onto the left-handed scalar and tensor operators in the WET, respectively. As these substantially mix under RGE, the fully leptonic $B\to \ell \nu$ processes are sensitive not only to the scalar coefficient but also to the tensor one. Note that the constraints from leptonic decays are notably stronger in the muon channel compared to the electron channel, as already anticipated from Section~\ref{sec:WET}. In order to close the flat direction appearing in the constraints from these processes, complementary constraints are needed. We overlay in both cases the constraints from exclusive semileptonic $b\to u \ell \nu$ transitions, as well as measurements of high-mass DY tails in the charged current channels. The former currently presents a better constraint in the considered scenarios, however, the latter is almost as important, especially for the tensor operator in the electron channel. The muon channel of high-mass DY exhibits a small tension with respect to the SM in this scenario, resulting in degenerate minima appearing in the fit and comparatively worsening the constraint. Lastly, we comment that complementary constraints from leptonic charm decays $D\to \ell \ell$ are currently not competitive when compared to fully leptonic $B\to \ell \nu$ transitions in these planes, however, the situation might change in the future with more data~\cite{Fajfer:2015mia}.

\begin{figure}[t]
     \centering
     \begin{subfigure}[b]{0.49\textwidth}
         \centering
         \includegraphics[width=\textwidth]{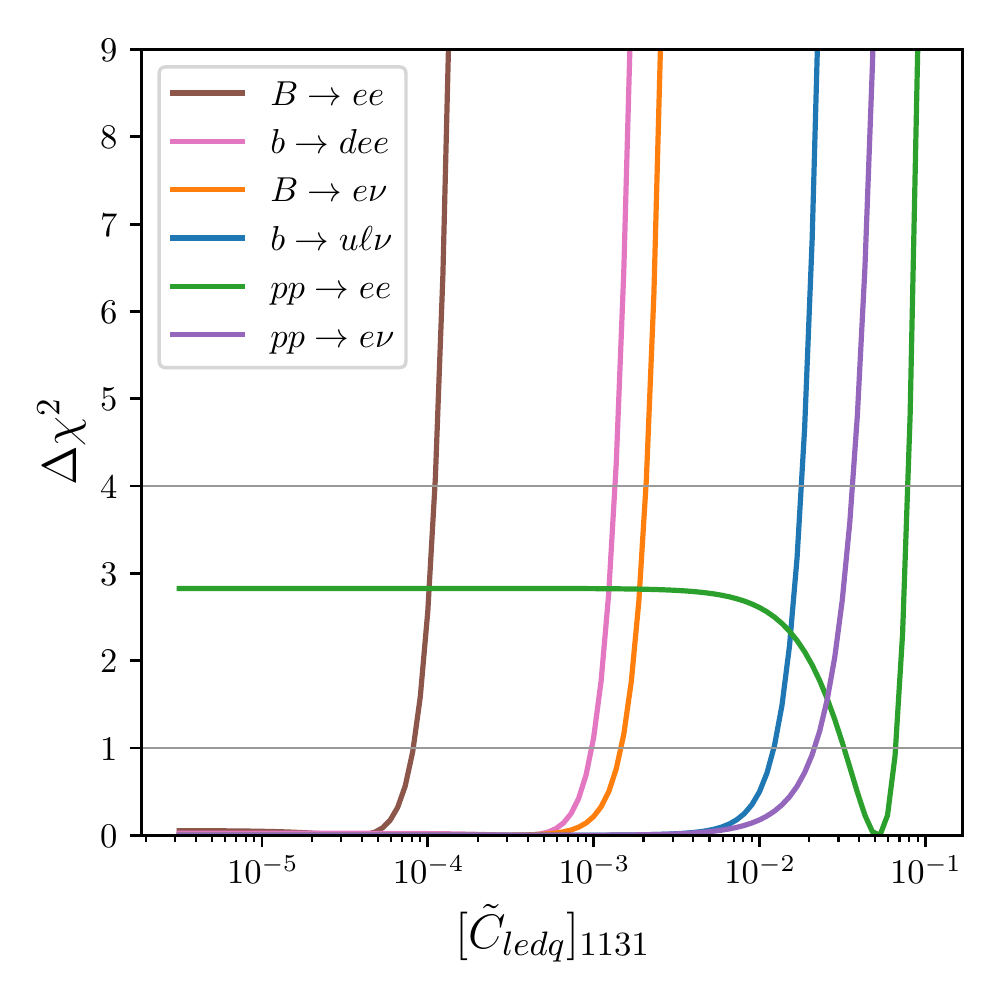}
     \end{subfigure}~
     \begin{subfigure}[b]{0.49\textwidth}
         \centering
         \includegraphics[width=\textwidth]{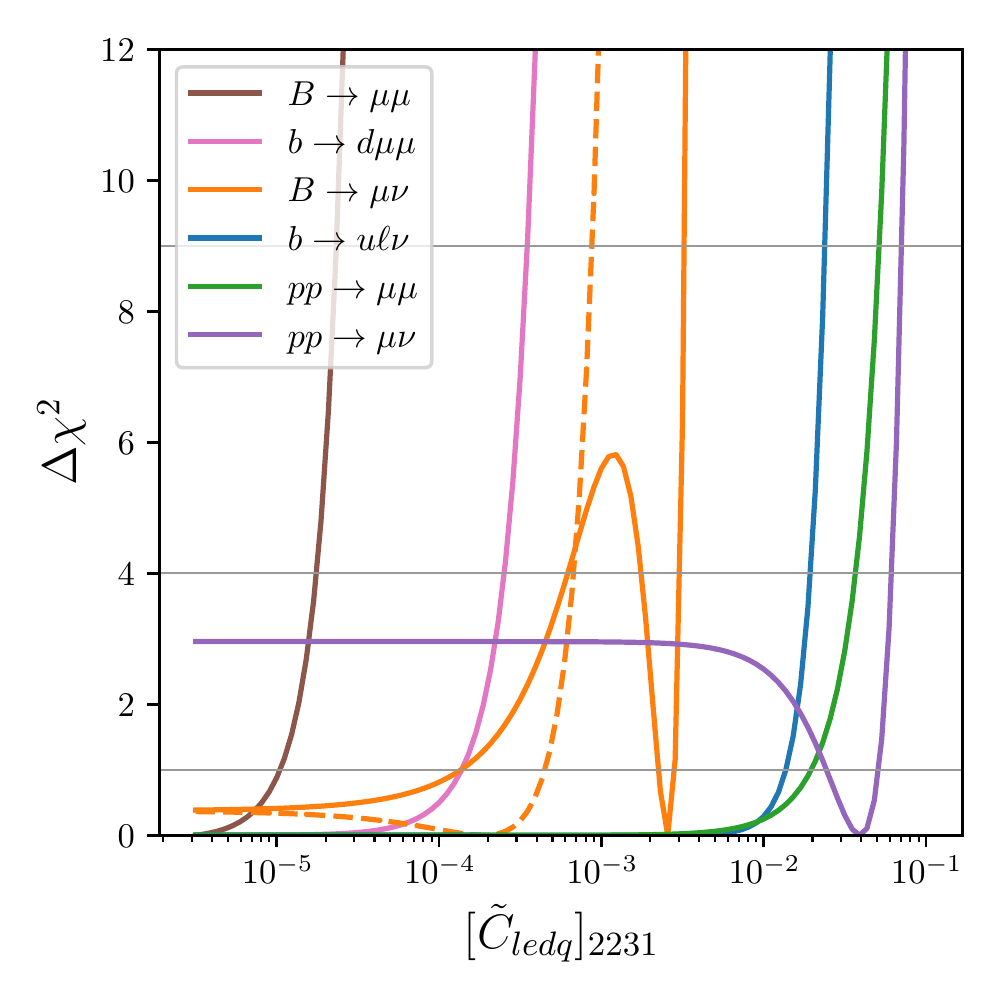}
     \end{subfigure}
        \caption{
        \textbf{Left:} $\Delta \chi^2$ in the scenario $[\tilde{C}_{ledq}]_{1131}$ (NP only in electrons).
        \textbf{Right:} $\Delta \chi^2$ in the scenario $[\tilde{C}_{ledq}]_{2231}$ (NP only in muons). See surrounding text for details.}
        \label{fig:smeft_ledq}
\end{figure}

Finally, in Figure~\ref{fig:smeft_ledq} we consider the $Q_{ledq}$ operator, assuming NP in either electrons (left) or muons (right). In these $1$D scenarios, we present the $\Delta\chi^2$ distributions, demonstrating the hierarchy between each of the relevant constraints. In both cases, the chirality-enhanced fully leptonic FCNC decays $B\to \ell \ell$ are the most efficient probes. They are followed by the FCNC transitions $b\to d\ell\ell$ which are comparable to chirality-enhanced charged current decays $B\to \ell\nu$. Note the difference between the solid and dashed orange lines in the right plot, corresponding to the positive and negative values of $\tilde{C}_{ledq}$, respectively. The second minimum is possible for the positive values of the WC due to cancellation with the SM, as already discussed in Section~\ref{sec:WET}. The rest of the bounds are symmetric with respect to the sign of the WCs. The exclusive $b\to u \ell \nu$ processes are last in sensitivity among the low-energy probes, followed by the complementary constraints from high-mass DY tails.

\section{Models}
\label{sec:models}

There is a finite number of heavy new field representations under the SM gauge group in a perturbative UV model that integrate out to the dimension-6 SMEFT operators at the tree level~\cite{deBlas:2017xtg}. Those contributions, that are leading both in the EFT and the loop expansion, are expected to dominate the phenomenology at low energies. In Subsection~\ref{sec:tree-models}, we map out all such mediators and their minimal set of couplings needed to generate a sizeable effect in $b \to u \ell \nu$ decays. Then, in Subsection~\ref{sec:example}, we elaborate on a specific example.

\subsection{Tree-level mediators}
\label{sec:tree-models}

We understand the SMEFT as a low-energy limit of a generic UV theory. Focusing on perturbative extensions of the SM, the natural next step after considering the SMEFT is to study the tree-level models that can be integrated out and generate the SMEFT operators at mass dimension $6$. It turns out there are only a finite number of scalar, fermion, and vector UV mediators one can consider under these conditions, as determined by the requirement of linear couplings to the SM fields~\cite{deBlas:2017xtg}. Moreover, the coupling structure of each mediator can impose non-trivial correlations in the SMEFT parameter space.

In this section, we explore such UV mediators, which can be integrated out in order to generate the operators interesting for $b\to u \ell \nu$ transitions, as collected in Table~\ref{tab:SMEFToperators} and discussed in detail in Section~\ref{sec:SMEFT}. We find that there are $4$ scalar, $5$ fermion, and $5$ vector degrees of freedom to which the exclusive semileptonic $b\to u \ell \nu$ decays can be in principle sensitive. We collect them in Table~\ref{tab:mediators}, where we follow the naming convention from Ref.~\cite{deBlas:2017xtg}, and provide their quantum numbers under the SM gauge group, as well as their spin. We group them by the operator, which they can generate so that each mediator can appear more than once in the table. 

\begin{table*}[t]
\center
\begin{tabular}{c|c}
\textbf{Operator}                                             & \textbf{Mediator}       \\ \hline
\multirow{5}{*}{$ [Q_{lq}^{(3)}]_{\ell \ell13}$}   & $\omega_1 \sim (\mathbf{3},\mathbf{1},-\frac{1}{3})_S$     \\
                                                     & $\zeta \sim (\mathbf{3},\mathbf{3},-\frac{1}{3})_S$        \\
                                                     & $\mathcal W \sim (\mathbf{1},\mathbf{3},0)_V$   \\
                                                     & $\mathcal U_2 \sim (\mathbf{3},\mathbf{1},\frac{2}{3})_V$ \\
                                                     & $\mathcal X  \sim (\mathbf{3},\mathbf{3},\frac{2}{3})_V$   \\ \hline
\multirow{5}{*}{$ [Q_{\phi q}^{(3)}]_{13}$} & $U \sim (\mathbf{3},\mathbf{1},\frac{2}{3})_F$            \\
                                                     & $D \sim (\mathbf{3},\mathbf{1},-\frac{1}{3})_F$            \\ 
                                                     & $T_1 \sim (\mathbf{3},\mathbf{3},-\frac{1}{3})_F$          \\ 
                                                     & $T_2 \sim (\mathbf{3},\mathbf{3},\frac{2}{3})_F$          \\
                                                     & $\mathcal W \sim (\mathbf{1},\mathbf{3},0)_V$   
\end{tabular}~~~~
\begin{tabular}{c|c}
\textbf{Operator}                                             & \textbf{Mediator}       \\ \hline
 \multirow{3}{*}{$ [Q_{ledq}]_{\ell \ell31}$}       & $\varphi \sim (\mathbf{1},\mathbf{2},\frac{1}{2})_S$      \\
                                                     & $\mathcal U_2 \sim (\mathbf{3},\mathbf{1},\frac{2}{3})_V$ \\
                                                    & $\mathcal Q_5 \sim (\mathbf{3},\mathbf{2},-\frac{5}{6})_V$ \\  \hline
\multirow{3}{*}{$ [Q_{lequ}^{(1)}]_{\ell \ell31}$} & $\varphi \sim (\mathbf{1},\mathbf{2},\frac{1}{2})_S$      \\
                                                      & $\omega_1 \sim (\mathbf{3},\mathbf{1},-\frac{1}{3})_S$     \\ 
                                                      & $\Pi_7 \sim (\mathbf{3},\mathbf{2},\frac{7}{6})_S$        \\ \hline
\multirow{2}{*}{$ [Q_{lequ}^{(3)}]_{\ell \ell31}$} & $\omega_1 \sim (\mathbf{3},\mathbf{1},-\frac{1}{3})_S$     \\
                                                     & $\Pi_7 \sim (\mathbf{3},\mathbf{2},\frac{7}{6})_S$        \\   \hline
\multirow{2}{*}{$ [Q_{\phi ud}]_{13}$}        & $Q_1 \sim (\mathbf{3},\mathbf{2},\frac{1}{6})_F$          \\
                                                     & $\mathcal B_1 \sim (\mathbf{1},\mathbf{1},1)_V$
\end{tabular}
\caption{List of all tree-level mediators generating operators contributing to exclusive $b\to u\ell \nu$ processes in the SMEFT at mass dimension $6$. The quantum numbers of the mediators are indicated as $(SU(3)_c,SU(2)_L,U(1)_Y)$ with the subscript denoting the spin of the mediator. }
\label{tab:mediators}
\end{table*}

The flavor structure of each mediator depends on its couplings with the SM fermions and is, in principle, completely free. One approach to lowering the number of free parameters is to endow them with a particular flavor assumption, see e.g.~\cite{Greljo:2023adz} for such a study under the MFV assumption. In this work, we take the minimalistic approach, in which we only consider the minimal set of couplings required to generate the correct flavor structure so as to contribute to $b\to u \ell \nu$ at the tree level. As a further simplification, we assume in all cases LFU in light leptons, as most of the data for $b\to u \ell \nu$ is reported as averages; see Section~\ref{sec:WET} for details. In the case of leptoquarks, this requires assuming them to be doublets under a leptonic $SU(2)$ flavor symmetry, which furthermore forbids lepton flavor violating operators, see~\cite{Greljo:2022jac} for a recent example. In the case of mediators which couple to leptons diagonally, only the leptons themselves have to be assumed to transform under a (diagonal) lepton flavor symmetry. As we will see in the following, even under this minimalistic approach, there are interesting correlations generated between different SMEFT operators and, moreover, between observables from different sectors. 

In Appendix~\ref{app:mediators}, we provide the minimal UV interaction Lagrangian of each mediator, and the corresponding tree-level matching onto the SMEFT WCs of mass dimension 6. We point out that each mediator requires two couplings to generate the operators interesting for $b\to u \ell \nu$ transitions. This, in turn, means that many of them generate further operators beyond those of initial interest, such as four-quark and four-lepton operators. In Subsection~\ref{sec:example}, we will explore the implications of this proliferation of operators in a concrete model example with two mediators, $\omega_1$ and $Q_1$.

\begin{figure}[t]
     \centering
     \begin{subfigure}[b]{0.49\textwidth}
         \centering
         \includegraphics[width=\textwidth]{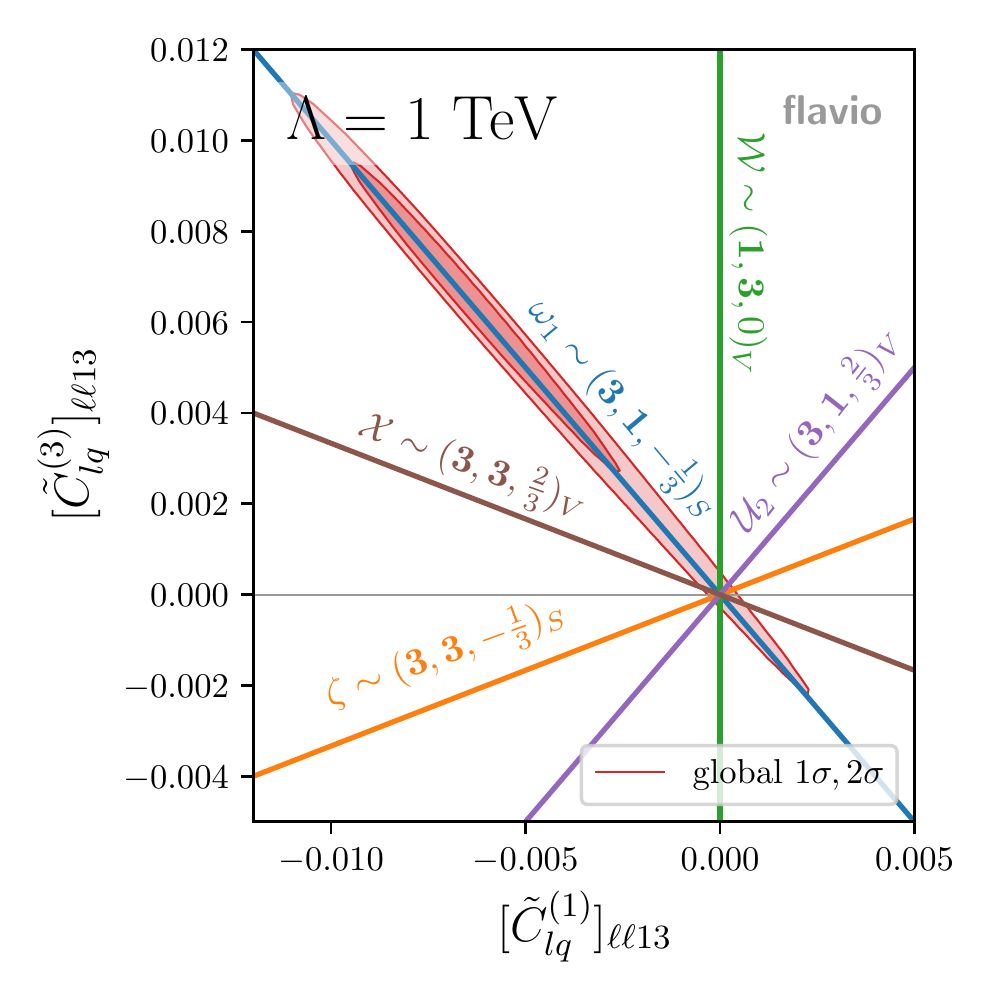}
     \end{subfigure}~
     \begin{subfigure}[b]{0.49\textwidth}
         \centering
         \includegraphics[width=\textwidth]{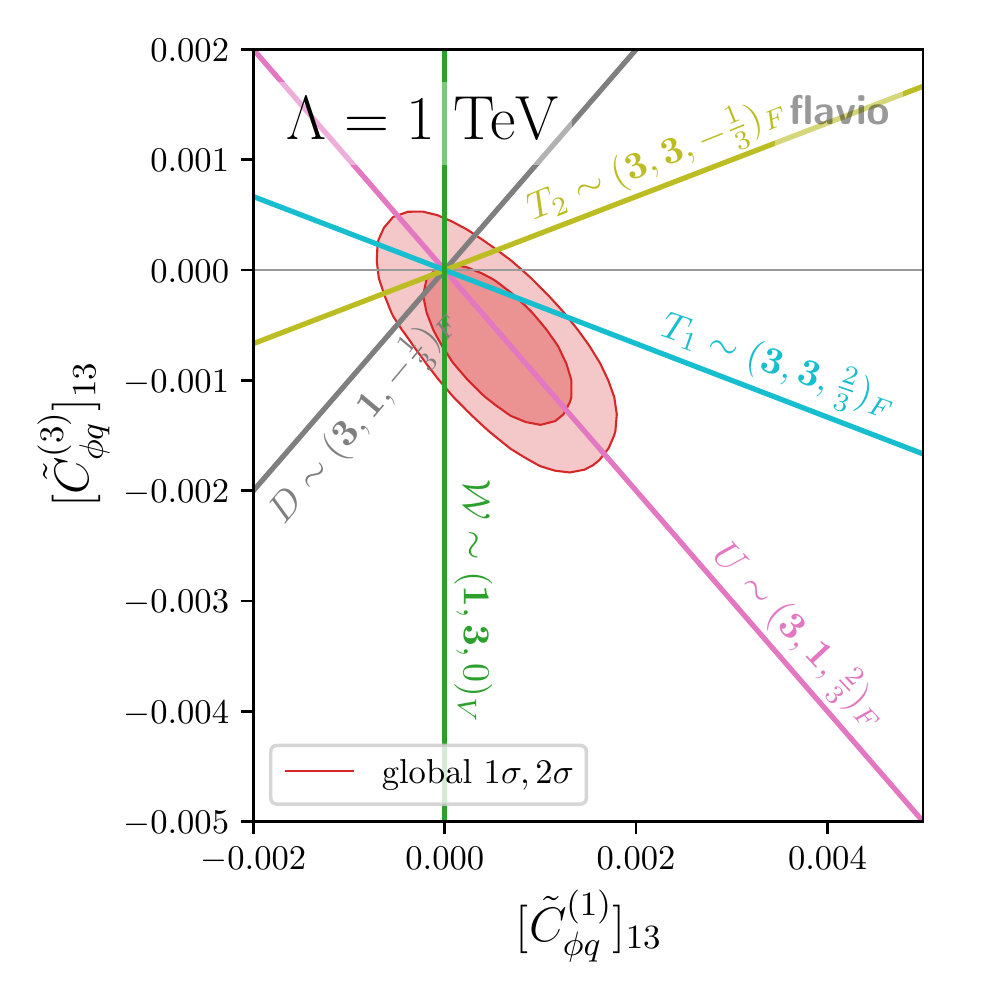}
     \end{subfigure}\\
     \begin{subfigure}[b]{0.49\textwidth}
         \centering
         \includegraphics[width=\textwidth]{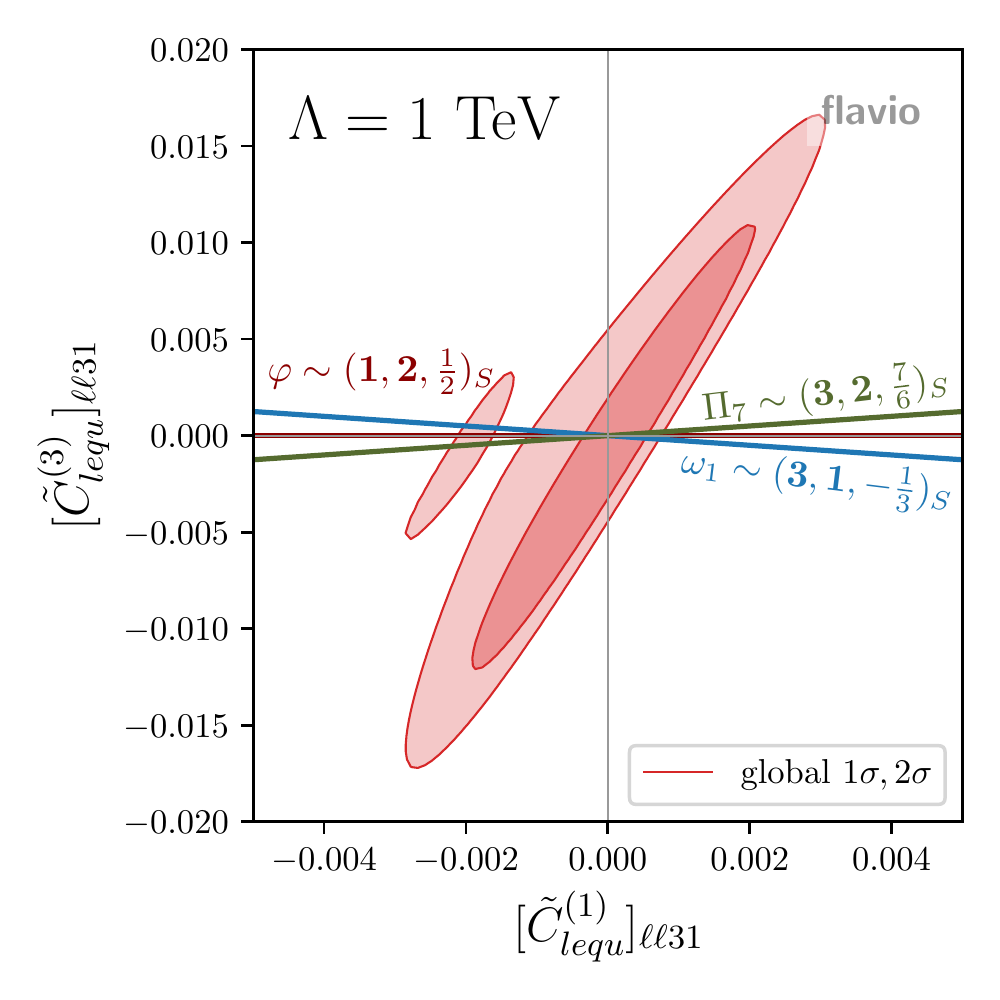}
     \end{subfigure}
        \caption{
        The $1\sigma$ and $2\sigma$ contours of a global SMEFT fit, overlaid with directions generated by single-mediator tree-level models in the scenarios 
        $([\tilde C_{lq}^{(1)}]_{\ell\ell 13}, [\tilde C_{lq}^{(3)}]_{\ell\ell 13})$, 
        $([\tilde C_{\phi q}^{(1)}]_{13}, [\tilde C_{\phi q}^{(3)}]_{13})$, and $([\tilde C_{lequ}^{(1)}]_{\ell\ell 31}, [\tilde C_{lequ}^{(3)}]_{\ell\ell 31})$. See text for detailed discussion.
        }
        \label{fig:1D_models}
\end{figure}

For the remainder of this subsection, we delve deeper into the correlations imposed by the UV mediators on the operators relevant to $b\to u \ell \nu$ processes. By studying the matching conditions in Appendix~\ref{app:mediators}, we find that interesting directions are generated by the UV mediators in the sets of WCs $(C_{lq}^{(1)},C_{lq}^{(3)})$, $(C_{\phi q}^{(1)},C_{\phi q}^{(3)})$, and $(C_{lequ}^{(1)},C_{lequ}^{(3)})$. In Figure~\ref{fig:1D_models}, we illustrate these directions in each of these planes, with each line corresponding to a correlation as implied by the interaction Lagrangian of each respective UV mediator. Moreover, we overlay the $1\sigma$ and $2\sigma$ contours as obtained from global SMEFT fits in Section~\ref{sec:SMEFT}. These are the same as in Figures \ref{fig:smeft_lq1_lq3_phiud} and \ref{fig:smeft_phiq1_phiq3_phiud}, whereas in the case of $(C_{lequ}^{(1)},C_{lequ}^{(3)})$ we redo the global fit from Figure~\ref{fig:smeft_lequ1_lequ3} assuming lepton flavor universality, in line with our assumptions about the UV mediators. 

In the upper left plot of Figure~\ref{fig:1D_models} we study the plane of $(\tilde{C}_{lq}^{(1)}, \tilde{C}_{lq}^{(3)})$. As discussed in Section~\ref{sec:SMEFT}, the direction of $\tilde{C}_{lq}^{(1)}=\tilde{C}_{lq}^{(3)}$ is tightly constrained from FCNC $b\to d \ell \ell$ processes. Only in the perpendicular direction of $\tilde{C}_{lq}^{(1)}=-\tilde{C}_{lq}^{(3)}$ do exclusive $b\to u \ell \nu$ processes play a significant role on the global fit. When interpreting the figure in terms of the directions in the SMEFT parameter space imposed by various UV mediators, the scalar leptoquark $\omega_1$ is singled out as the only viable mediator which can be probed by exclusive $b\to u \ell \nu$ processes. All of the other UV mediators, which are shown on the plot, and which, in principle, generate operators that could contribute to these processes, are constrained significantly more through complementary measurements. Combining several mediators at the same time would allow for cancellations of new physics contributions in $b \to d \ell 
\ell$, which would consequently align with the SM prediction. However, we find this scenario less desirable from our standpoint.

A similar conclusion can be made when considering the $(\tilde{C}_{\phi q}^{(1)}, \tilde{C}_{\phi q}^{(3)})$ plane in the upper right plot of Figure~\ref{fig:1D_models}. In this case, the global fit is not dominated by the exclusive $b\to u \ell \nu$, as can be seen in Figure~\ref{fig:smeft_phiq1_phiq3_phiud} left by comparing the global fit with the blue region. The least constrained single-mediator extension, not prone to significant constraints from complementary measurements, is the vectorlike fermion $U$, as the direction it implies aligns with the flat direction in stringent $b\to d \ell \ell$ processes, up to small RG corrections, as discussed in Section~\ref{sec:SMEFT}. Given the complementary bounds from neutral meson mixing, even this mediator can not substantially contribute to exclusive $b \to u \ell \nu$ decays. 

The bottom panel of Figure~\ref{fig:1D_models} shows that all the mediators contributing to $(\tilde{C}_{lequ}^{(1)}, \tilde{C}_{lequ}^{(3)})$ generate directions which are best constrained from fully leptonic $B\to \ell \nu$ transitions, and hence the exclusive semileptonic $b\to u \ell \nu$ transitions are inefficient probes of their effects. Similarly, the mediators from Table~\ref{tab:mediators}, generating $Q_{ledq}$, will be constrained from the fully leptonic decays; see Eq.~\eqref{eq:CT}, Table~\ref{tab:WET-1D-bounds}, and Figure~\ref{fig:smeft_ledq}.

Finally, we acknowledge again that the above conclusions come under the assumption of a single UV mediator. Although this assumption might be unrealistic, it is clear that cancellations tuned to keep observables SM-like in a model of multiple mediators would be needed in order to make exclusive $b\to u \ell \nu$ decays relevant. Barring such cancellations, we demonstrated that these decays could act as potential probes of only a handful of UV degrees of freedom that integrate out onto the dimension~6 SMEFT at the tree level. Other than $\omega_1$ already pointed out in the text so far, $Q_1$ and $\mathcal{B}_1$ are the only mediators since they match onto the operator $Q_{\phi u d}$.

\subsection{Explicit example: $\omega_1$ and $Q_1$}
\label{sec:example}

As a final example, consider extending the SM with two heavy fields: a scalar leptoquark $\omega_1 \sim (\mathbf{3},\mathbf{1},-\frac{1}{3})_S$ of mass $M_{\omega_1}$ and a vector-like partner of the left-handed quark doublet $Q_1 \sim (\mathbf{3},\mathbf{2},\frac{1}{6})_F$ of mass $M_{Q_1}$. For simplicity, we focus only on a minimal set of couplings required to produce effects in semileptonic $b\to u \ell \nu$ transitions and investigate the consequences from complementary constraints (see e.g.~\cite{Crivellin:2022rhw} for a recent study of VLQs coupling to first and second generation quarks).

We focus on the interesting scenario, in which $\omega_1$ is responsible for generating the operator $[Q_{lq}^{(3)}]_{\ell \ell 13} = -[Q_{lq}^{(1)}]_{\ell \ell 13}$ and $Q_1$ generates the operator $[Q_{\phi ud}]_{13}$. The UV interaction Lagrangians for this scenario are given in Eq.~\eqref{eq:UVLagrangian_omega1_lq3} for the leptoquark and in Eq.~\eqref{eq:UVLagrangian_Q1} for the vector-like quark. There are no terms in the UV Lagrangian involving both $\omega_1$ and $Q_1$ simultaneously that would match onto SMEFT operators of dimension 6 at the tree level. Both $\omega_1$ and $Q_1$ have two couplings of interest: $(y_{\omega_1}^{ql})_{1}$ and $(y_{\omega_1}^{ql})_{3}$, denoting the couplings of the leptoquark to the left-handed first and third generation quark doublets and light leptons (universally), and $(\lambda_{Q_1}^{u})_{1}$ and $(\lambda_{Q_1}^{d})_{3}$, denoting the Yukawa couplings between the vector-like quark, the Higgs, and the first generation right-handed up quarks and third generation right-handed down quarks. This set of couplings is however not only responsible for generating the set of operators required for $b\to u \ell \nu$ transitions, but also a set of other SMEFT operators. For completeness, we list here the full matching conditions of the model onto the set of generated SMEFT operators:\footnote{We omit here the operators $ [Q_{u\phi}]_{i1}$ and $ [Q_{d\phi}]_{i3}$ that modify the quark interactions with the Higgs boson, as they are not expected to pose any competitive bounds.}
\begin{align}
\left[C_{lq}^{(1)}\right]_{\ell\ell13} &= - \left[C_{lq}^{(3)}\right]_{\ell\ell13} = \frac{(y^{ql}_{\omega_1})_{1}^*(y^{ql}_{\omega_1})_{3}}{4M_{\omega_1}^2} \,,\\
\left[C_{lq}^{(1)}\right]_{\ell\ell11} &= - \left[C_{lq}^{(3)}\right]_{\ell\ell11} = \frac{|(y^{ql}_{\omega_1})_{1}|^2}{4M_{\omega_1}^2} \,,\\
\left[C_{lq}^{(1)}\right]_{\ell\ell33} &= - \left[C_{lq}^{(3)}\right]_{\ell\ell33} = \frac{|(y^{ql}_{\omega_1})_{3}|^2}{4M_{\omega_1}^2} \,,\\
\left[C_{\phi ud}\right]_{13} &= \frac{(\lambda^d_{Q_1})_{3}(\lambda^u_{Q_1})_{1}^*}{M_{Q_1}^2} \,, \\
\left[C_{\phi u}\right]_{11} &= -\frac{|(\lambda^u_{Q_1})_{1}|^2}{2M_{Q_1}^2} \,, \\
\left[C_{\phi d}\right]_{33} &= \frac{|(\lambda^d_{Q_1})_{3}|^2}{2M_{Q_1}^2}\,.
\end{align}
In order for the WCs to carry a sufficient complex phase, in line with the assumption already presented in Eq.~\eqref{eq:SMEFTredefine}, we assume that one of the couplings for each mediator is complex, and define
\begin{align}
    (y^{ql}_{\omega_1})_{3} = \frac{V_{ub}}{|V_{ub}|}(\tilde{y}^{ql}_{\omega_1})_{3} \,,\\
    (\lambda^d_{Q_1})_{3} = \frac{V_{ub}}{|V_{ub}|}(\tilde{\lambda}^d_{Q_1})_{3} \,,
\end{align}
where $\tilde{y}$ and $\tilde{\lambda}$ are real parameters. Note, that this model does not contribute significantly to the loop processes that can be used to extract the CKM parameters~\cite{Altmannshofer:2021uub, Brod:2019rzc}. We therefore fix $|V_{ub}|$ as elsewhere in the main part of this paper, which is consistent with Eq.~(47) of~\cite{Altmannshofer:2021uub}.

\begin{figure}[t]
     \centering
     \begin{subfigure}[b]{0.8\textwidth}
         \centering
         \includegraphics[width=\textwidth]{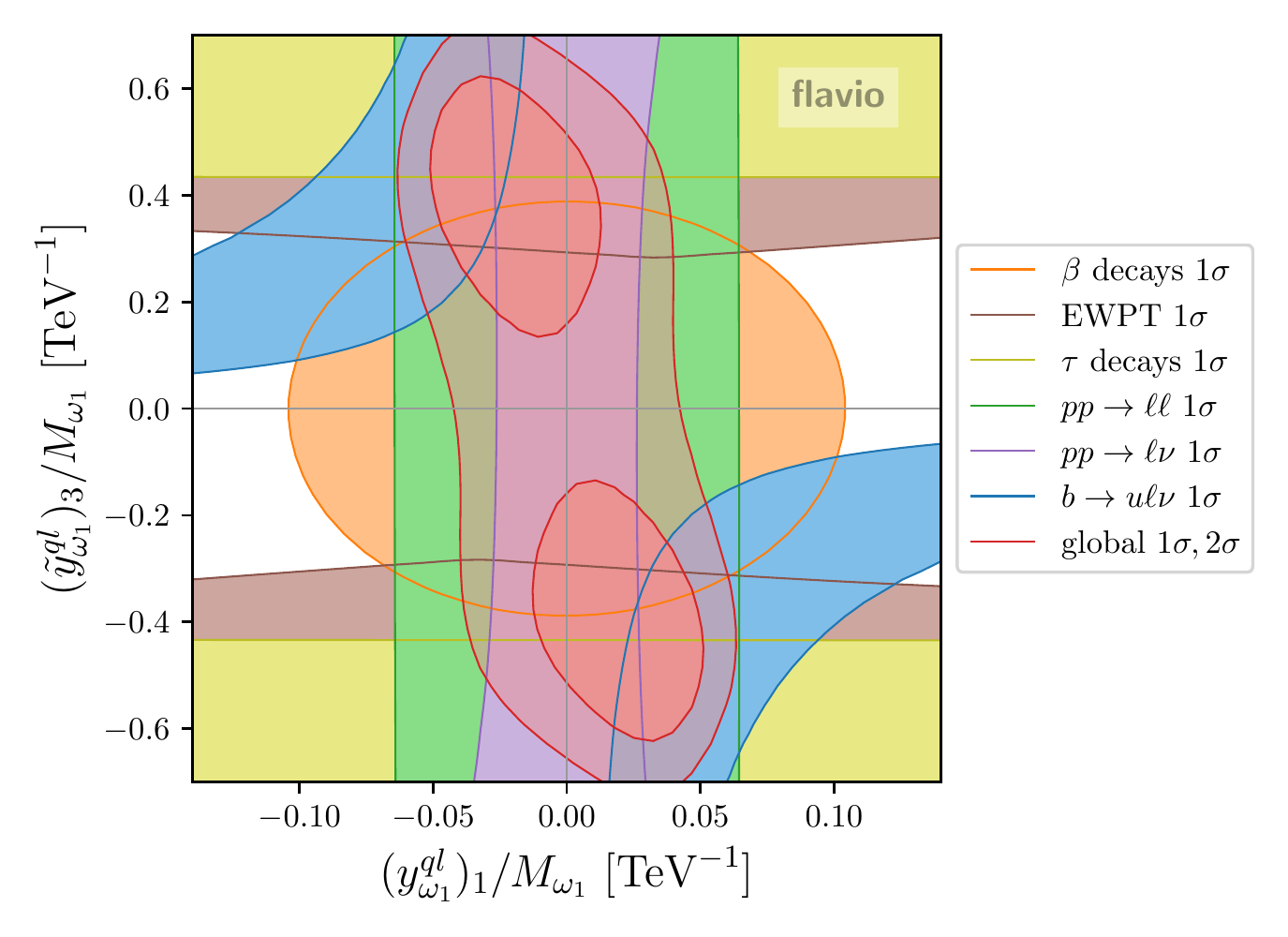}
     \end{subfigure}~
        \caption{Fit to the data in the explicit model example ($\omega_1$ plus $Q_1$). For details see Section~\ref{sec:example}.}
        \label{fig:model_w1}
\end{figure}

We perform a study of the $4$-dimensional model parameter space using the same numerical setup as described in Sections \ref{sec:WET} and \ref{sec:SMEFT}. For presentation purposes, we decide to show the results in the plane of the LQ couplings $((y^{ql}_{\omega_1})_{1}/M_{\omega_1}, (\tilde{y}^{ql}_{\omega_1})_{3}/M_{\omega_1})$ by profiling the global likelihood over the two parameters of the vector-like quark $Q_1$. The individual likelihoods of the relevant processes are then evaluated using the same parameter values as for the global likelihood. The results are shown in Figure~\ref{fig:model_w1}. 

A few comments are in order, firstly, the global fit of the model marginally improves the global $\chi^2$ by $\Delta \chi^2=-3.72$ with respect to the SM. Moreover, there are many complementary constraints participating in constraining the model parameter space. The $b\to u \ell \nu$ processes show a slight tension with the SM point, as discussed already in Sections~\ref{sec:WET} and \ref{sec:SMEFT}. However, notice that the high-mass DY tails, especially the charged current processes, provide severe constraints on the direction in which the LQ couples to valence quarks. Both directions in Figure~\ref{fig:model_w1} are constrained also from measurements of super-allowed $\beta$-decays~\cite{Hardy:2020qwl, Falkowski:2020pma}, either at the tree level (the $[Q_{lq}^{(3)}]_{\ell\ell11}$ operator), or through RG effects ($[Q_{lq}^{(3)}]_{\ell\ell33}$ mixing into $[Q_{lq}^{(3)}]_{\ell\ell11}$). Lastly, the EWPT and leptonic $\tau$ decay constraints, both showing slight tensions in participating observables (see Appendix in Ref.~\cite{Aebischer:2018iyb}), contribute to the pull of the global best fit region away from the SM point. Both $\omega_1$ and $Q_1$ can contribute to these processes, either through modified $Z$ couplings with right-handed quarks ($Q_1$), or via further RGE-induced contributions ($\omega_1$)~\cite{Feruglio:2017rjo}.

It is instructive to contrast our findings with the direct search limits emerging from the LHC. These limits are predominantly derived from the QCD-induced pair production of scalar leptoquarks, which decay chiefly to third-generation quarks and light leptons, with $M_{\omega_1}\gtrsim 1.4$\,TeV~\cite{ATLAS:2020xov}. In summary, while the complementary constraints diminish the significance of the $B \to V \ell \nu$ tension, it remains noteworthy that there exists a parameter space where a leptoquark, consistent with direct search results and exhibiting $\mathcal{O}(1)$ couplings to third-generation quarks and substantially smaller couplings with light quarks, renders the exclusive $b \to u \ell \nu$ decays relevant. Lastly, we note that both of the vector-like quark couplings are perturbative in the whole plot range of Figure~\ref{fig:model_w1}. At the best fit point, they take the values of $((\lambda^u_{Q_1})_{3}/M_{Q_1},(\tilde{\lambda}^d_{Q_1})_{3}/M_{Q_1}) = (0.13, 0.08)~ \mathrm{TeV}^{-1}$, comfortably allowing for a VLQ of mass $M_{Q_1}\gtrsim 1.2\,\mathrm{TeV}$ with perturbative couplings, in compliance with direct searches~\cite{CMS:2018wpl}.

\section{Conclusions}
\label{sec:conc}

The decays of $B$ hadrons provide an exceptional environment for probing the intricacies of particle physics. This research program is notably promising, given the breadth of experimental activity currently underway and anticipated for this decade. An enhanced understanding of strong dynamics further bolsters it, attributable to contemporary advancements in lattice QCD. The ultimate goal of this precision frontier is to examine the SM and possibly uncover NP effects rigorously. The pressing question arises: what kind of new physics can we anticipate to investigate? The bottom-up EFT approach provides us with a systematic framework for addressing this question for an arbitrary short-distance new physics.

This work investigates the NP potential of exclusive $b \to u \ell \nu$ decays, commonly used to extract the SM input parameter $|V_{ub}|$. We start out in Section~\ref{sec:WET}, where we perform a comprehensive analysis in the context of the WET within the \texttt{flavio} framework, building the likelihood from the experimental data and theoretical predictions, and deriving the optimal parameter space for the WCs. Our primary findings are encapsulated in Figures~\ref{fig:wet_CVLCVR} and \ref{fig:wet_CSLCSR}, and in Table~\ref{tab:WET-1D-bounds}. Most operators are constrained by semileptonic decays, with the exception of the pseudoscalar operator, which is dominantly restricted by the fully leptonic decay. While many preferred regions encompass the SM prediction, the axial vector operator slightly prefers a non-zero value due to the tension in $B \to V \ell \nu$ channel where $V=(\rho,\omega)$. To understand the role of $|V_{ub}|$ in the presence of such NP, we fit $|V_{ub}|$ concurrently with the axial vector operator, as illustrated in Figure~\ref{fig:wet_VubCVLCVR}.

What are the implications of the WET analysis on short-distance physics? To shed light on this query, we advance our examination by employing the SMEFT in Section~\ref{sec:SMEFT}. This framework inevitably predicts significant correlations with other processes. Our objective within this context was to conduct comprehensive SMEFT fits, presuming a complete set of dimension-6 operators at a high-energy scale (set to $\Lambda = 1 $\,TeV), which offer substantial contributions to exclusive $b \to u \ell \nu$ decays. This set is augmented by closely related operators that aid in attenuating the complementary bounds or that necessarily come along in a tree-level UV completion. The predictions of the SMEFT, intrinsically caused by the $SU(2)_L$ gauge symmetry and renormalization group evolution down to low energies, impact rare natural-current $b$ decays, $B^0 - \bar B^0$ mixing, high-mass Drell-Yan production, and $W/Z$ vertex corrections. These influences collectively drive the global fit results, succinctly presented in Figures~\ref{fig:smeft_lq1_lq3_phiud}, \ref{fig:smeft_phiq1_phiq3_phiud}, \ref{fig:smeft_lequ1_lequ3} and \ref{fig:smeft_ledq}. The conclusive finding of this examination indicates that exclusive $b \to u \ell \nu$ decays are instrumental in probing the set of operators defined as $([Q_{lq}^{(1)}]_{\ell\ell 13}=-[Q_{lq}^{(3)}]_{\ell\ell 13}, [Q_{\phi ud}]_{13})$, a result that is depicted in Figure~\ref{fig:smeft_lq1_lq3_phiud} (right).

Perturbative UV completions that match onto the dimension-6 SMEFT operators at the tree level can be systematically classified based on the gauge and Lorentz representation of the mediating heavy fields. A comprehensive list of all such cases that contribute at the leading order to exclusive $b \to u \ell \nu$ decays is presented in Table~\ref{tab:mediators} and expanded upon in Appendix~\ref{app:mediators}. A particularly noteworthy result is visually encapsulated in Figure~\ref{fig:1D_models}, elegantly portraying the correlations imposed by diverse heavy field mediators. Upon close scrutiny, it can be concluded that, absent any conspiring cancellations maintaining SM-like observables, promising cases are represented by $Q_1$ and $\mathcal{B}_1$, which correspond to the operator $Q_{\phi u d}$, along with $\omega_1$ generating the operator $Q_{lq}^{(1)}=-Q_{lq}^{(3)}$. In Subsection~\ref{sec:example}, we undertake an in-depth analysis of two such mediators simultaneously present, constraining our focus to a minimal set of couplings essential for the $b \to u \ell \nu$ decay process; see Figure~\ref{fig:model_w1}. The apparent difficulty in addressing the $B \to V \ell \nu$ tension, given the complementary constraints, underscores the necessity for further advancements in lattice QCD computations. In this regard, pioneering efforts, such as those represented by the work detailed in~\cite{Leskovec:2022ubd}, are of substantial importance and provide a promising path forward to resolving this puzzle.

Future improvements of this study could involve discarding the assumption of lepton flavor universality in exclusive $b \to u \ell \nu$ decays, a constraint currently necessitated by the scope of available data. We fervently advocate for experimental collaborations to conduct measurements of theoretically pristine $\mu/e$ ratios, in analogy with the proposals~\cite{Hiller:2003js, Isidori:2020eyd, Bordone:2021olx, Ardu:2022glb}. Additionally, it could be intriguing to go further than the current scope, exploring beyond tree-level matching and beyond the dimension-6 operators. However, we anticipate this would lead to tighter constraints. This is because any new heavy mediators brought into the picture would need to either be lighter or interact more strongly in order to make up for the effects of suppression. Finally, investigating correlations with the $b \to c \ell \nu$ decays, predicated by several motivated NP flavor structures within the quark sector~\cite{Faroughy:2020ina, Greljo:2022cah}, could significantly enhance our understanding of the relevance of these decays in the pursuit of physics beyond the SM.

\section*{Acknowledgements}

We thank Méril Reboud for the useful comparison with~\cite{Leljak:2023gna}. This work received funding from the Swiss National Science Foundation (SNF) through the Eccellenza Professorial Fellowship ``Flavor Physics at the High Energy Frontier'' project number 186866. AG is also partially supported by the European Research Council (ERC) under the European Union’s Horizon 2020 research and innovation program, grant agreement 833280 (FLAY).

\appendix

\section{$|V_{ub}|$ determination}
\label{app:Vub}

\begin{figure}[h!]
     \centering
     \begin{subfigure}[b]{0.49\textwidth}
         \centering
         \includegraphics[width=\textwidth]{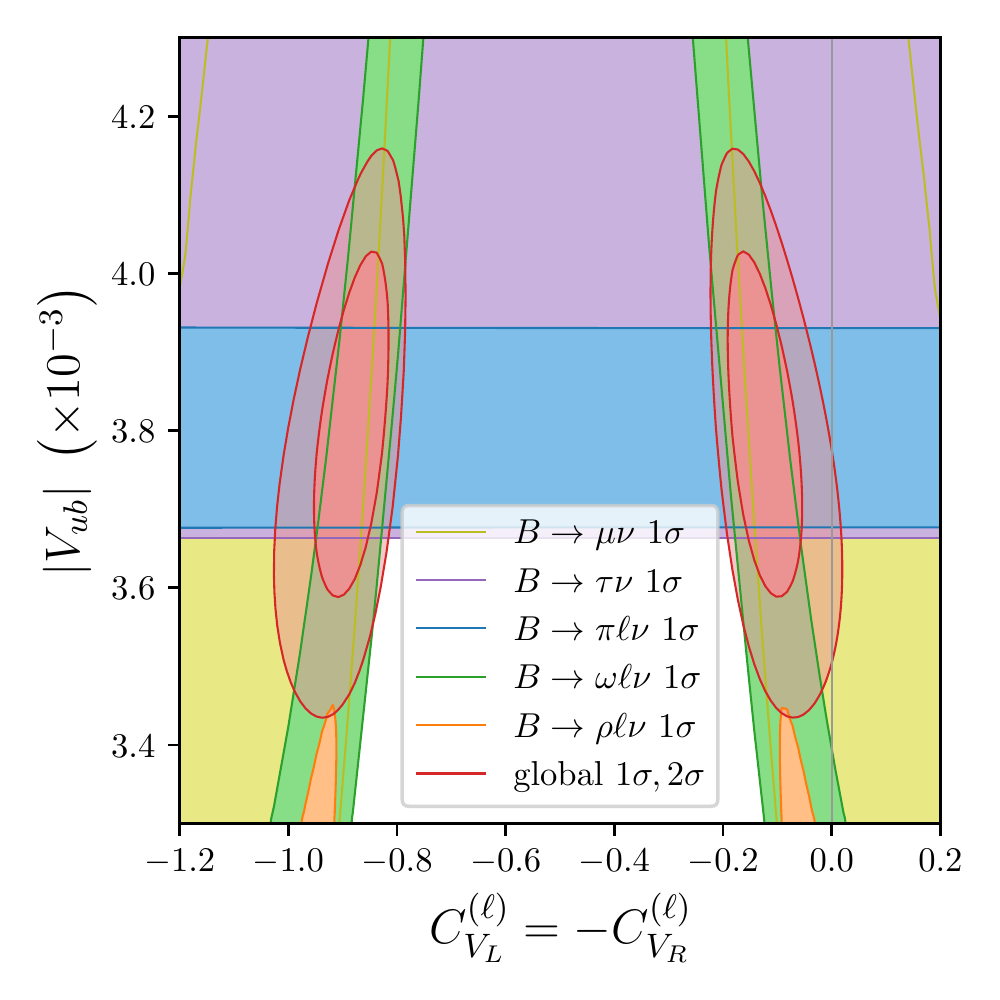}
     \end{subfigure}~
      \begin{subfigure}[b]{0.49\textwidth}
         \centering
         \includegraphics[width=\textwidth]{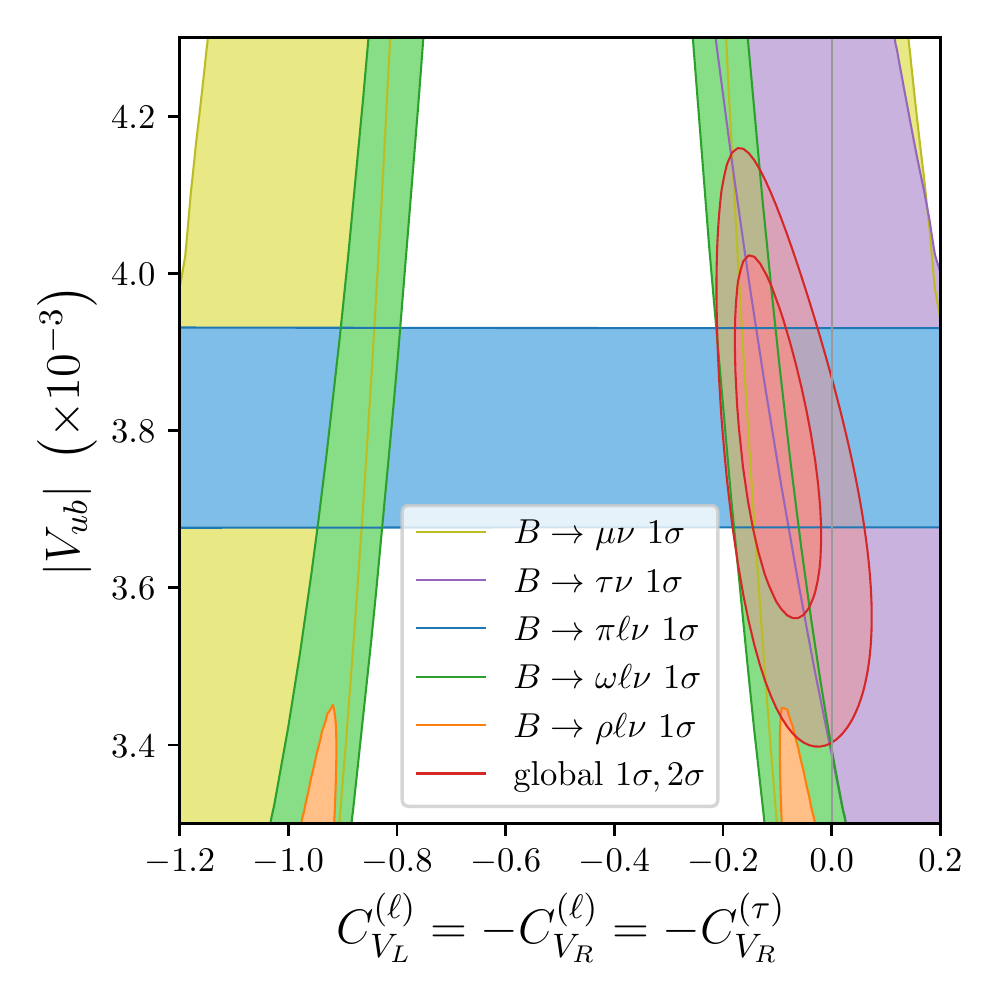}
     \end{subfigure}\\
           \begin{subfigure}[b]{0.49\textwidth}
         \centering
         \includegraphics[width=\textwidth]{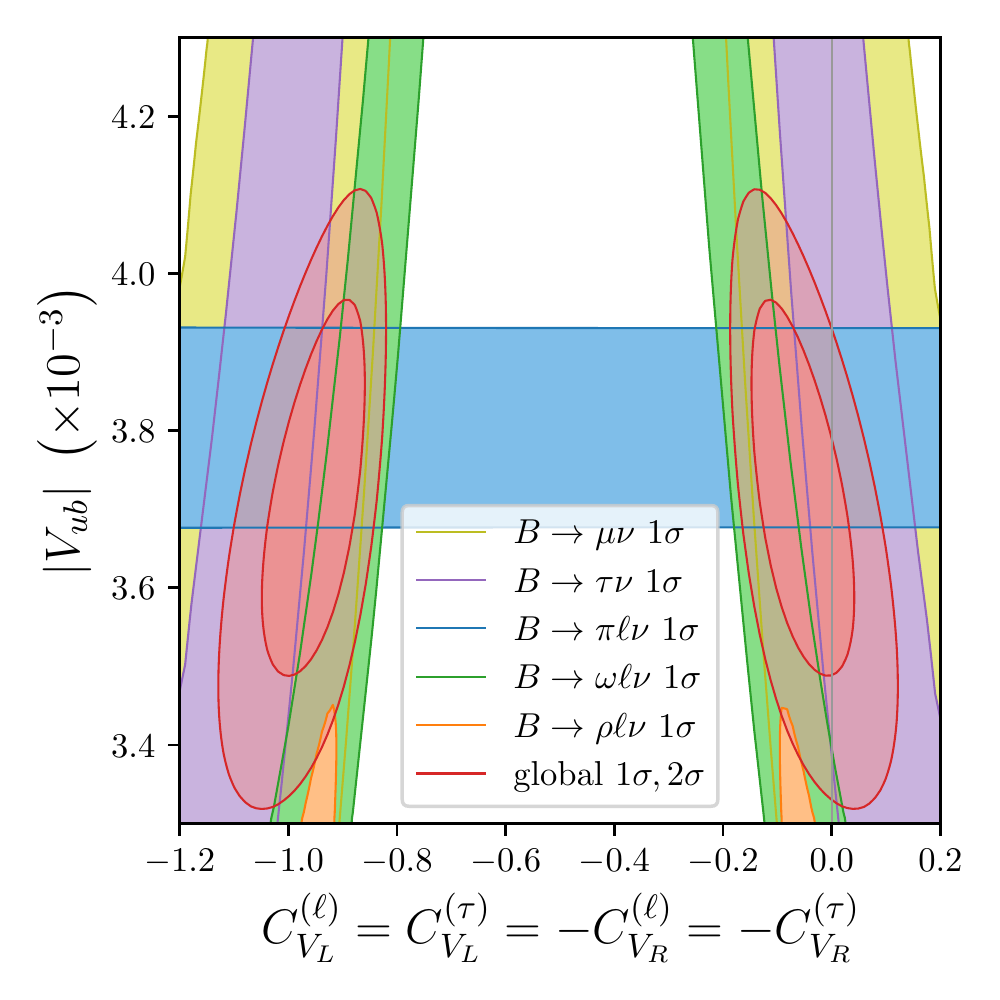}
     \end{subfigure}
        \caption{A combined determination of $|V_{ub}|$ and the axial-vector WET WC under three scenarios, as motivated by the SMEFT. \textbf{Left:} $|V_{ub}|$ and $C_{V_L}^{(\ell)}=-C_{V_R}^{(\ell)}$.
        \textbf{Right:} $|V_{ub}|$ and $C_{V_L}^{(\ell)}=-C_{V_R}^{(\ell)}$ assuming LFU in $C_{V_R}^{(\tau)} = C_{V_R}^{(\ell)}$. \textbf{Lower:} $|V_{ub}|$ and $C_{V_L}=-C_{V_R}$ assuming LFU for all leptons.}
        \label{fig:wet_VubCVLCVR}
\end{figure}

In this Appendix, we discuss the issue of extracting the CKM matrix parameter $|V_{ub}|$ together with potential short-distance NP effects, in light of the tension between $B\to \pi \ell \nu$ and $B\to V \ell \nu$ with $V=(\rho, \omega)$, as discussed in Section~\ref{sec:WET}. In Figure~\ref{fig:wet_CVLCVR} we have shown that the tension appears in the axial-vector direction of the WET WC parameter space, by assuming no NP contributions to $|V_{ub}|$. Here we argue that this is a valid assumption.

In Figure~\ref{fig:wet_VubCVLCVR} we perform a combined fit of $|V_{ub}|$ and the axial-vector WET operator, under three scenarios, as motivated by the SMEFT (see Section~\ref{sec:SMEFT}). We consider the constraints in these planes from $B\to \pi \ell \nu$, $B\to V \ell \nu$, as well as fully leptonic modes $B\to \mu \nu$ and $B\to \tau \nu$. The three scenarios presented on Figure~\ref{fig:wet_VubCVLCVR} correspond to the assumptions of no NP in the $\tau$ WCs (left plot), universal contributions to all leptons only in the right-handed vector WC (right plot), or universal contributions in both the left- and right-handed vector WCs (bottom plot). Note, that in all three cases $B\to \pi \ell \nu$ is by far the most constraining process in the $|V_{ub}|$ direction. The global fits, although changing slightly between the three scenarios, remain dominated by $B\to \pi \ell \nu$, which is completely insensitive to the axial-vector WC direction. We conclude that the phenomenological analysis presented in this paper, assuming the value of $|V_{ub}|$ consistent with the global fits using $B\to \pi \ell \nu$ only, is a valid and reasonable approach.

Lastly, in the case of vector, scalar and tensor WCs, the bounds on which are discussed both in the WET in Section~\ref{sec:WET}, $B \to \pi \ell \nu$ can play an important role in constraining the NP effects. In the scope of our paper, we can assume that in these cases the CKM matrix parameters are fixed by $\Delta F=2$ processes when those are unpolluted by NP effects. Of course, for a global SMEFT analysis, one should be cautious and perform a global fit of the SM input parameters together with the WCs, as argued in Ref.~\cite{Descotes-Genon:2018foz}.

\section{Lagrangians for the tree-level mediators}
\label{app:mediators}

This Appendix presents the single mediator Lagrangians for the tree-level models listed in Table \ref{tab:mediators}. 
We focus on cases where only the minimal set of non-zero couplings is present, resulting in the generation of a single effective operator contributing to $b\to u\ell\nu$. 
The Lagrangians considered here are restricted to renormalizable terms, with the kinetic and mass terms omitted for brevity.

Additionally, we provide the SMEFT Lagrangians obtained after integrating out the heavy mediators. 
It is important to note that each SMEFT Lagrangian implicitly includes the appropriate Hermitian conjugate terms to ensure that the final Lagrangian takes the form of Eq.~\eqref{eq:SMEFT}. 
The Lagrangians and the matching to the SMEFT framework are based on the work presented in \cite{deBlas:2017xtg}.

Furthermore, we note that the effective operators associated with lepton flavor indices exhibit lepton flavor universality for electrons and muons, as discussed in Section \ref{sec:models}. 
Finally, the notation $C_{ijkl} Q_{ijkl} + (ijkk + ijll)$ indicates that the Lagrangian also includes terms with the flavor indices specified within the parentheses, with appropriate substitutions made for both the operator and the WC.

\subsection{Scalars}
\subsubsection*{$\varphi \sim (\mathbf{1},\mathbf{2},\frac{1}{2})_S$}

\begin{itemize}
    \item $Q_{ledq}$
    \begin{itemize}
        \item UV Lagrangian:
    
    \begin{equation}
        -\mathcal L^{\leq 4} \supset (y^{d}_{\varphi})_{31}~ \varphi^{\dagger} \bar d_{3} q_{1} + (y^{e}_{\varphi})~ \varphi^{\dagger} \bar e_{\ell} l_{\ell} + \mathrm{h.c.}
    \end{equation}
    
        \item Matching to SMEFT:
    
    \begin{equation}
    \begin{split}
        \mathcal L_{\mathrm{SMEFT}} \supset & \frac{(y^d_\varphi)_{31}(y^e_\varphi)^*}{M_{\varphi}^2}[Q_{ledq}]_{\ell\ell31} \\
        & - \frac{|(y^d_\varphi)_{31}|^2}{6M_{\varphi}^2}\left([Q_{qd}^{(1)}]_{1133} + 6[Q_{qd}^{(8)}]_{1133}\right) - \frac{|(y^e_\varphi)|^2}{2M_{\varphi}^2}[Q_{le}]_{\ell\ell'\ell'\ell}
        \end{split}
    \end{equation}
    
    \end{itemize}
    \item $Q_{lequ}^{(1)}$
    \begin{itemize}
        \item UV Lagrangian:
    
    \begin{equation}
        -\mathcal L^{\leq 4} \supset (y^{u}_{\varphi})_{31}~ \varphi^{\dagger} i\sigma_2 \bar q_{3}^T u_{1} + (y^{e}_{\varphi})~ \varphi^{\dagger} \bar e_{\ell} l_{\ell} + \mathrm{h.c.}
    \end{equation}
    
        \item Matching to SMEFT:
    
    \begin{equation}
    \begin{split}
        \mathcal L_{\mathrm{SMEFT}} \supset & \frac{(y^u_\varphi)_{31}(y^e_\varphi)^*_{\ell\ell}}{M_{\varphi}^2}[Q_{lequ}^{(1)}]_{\ell\ell31} \\
        & - \frac{|(y^u_\varphi)_{31}|^2}{6M_{\varphi}^2}\left([Q_{qu}^{(1)}]_{3311} + 6[Q_{qu}^{(8)}]_{3311}\right) - \frac{|(y^e_\varphi)|^2}{2M_{\varphi}^2}[Q_{le}]_{\ell\ell'\ell'\ell}
        \end{split}
    \end{equation}
    
    \end{itemize}

\end{itemize}

\subsubsection*{$\omega_1 \sim (\mathbf{3},\mathbf{1},-\frac{1}{3})_S$}

\begin{itemize}
    \item $Q_{lq}^{(3)}$
    \begin{itemize}
        \item UV Lagrangian:
    
    \begin{equation}
        -\mathcal L^{\leq 4} \supset (y^{ql}_{\omega_1})_{1}~ \omega_{1\ell}^\dagger \bar q_{1}^c i\sigma_2 l_{\ell} + (y^{ql}_{\omega_{1}})_{3}~ \omega_{1\ell}^\dagger \bar q_{3}^c i\sigma_2 l_{\ell} + \mathrm{h.c.}
        \label{eq:UVLagrangian_omega1_lq3}
    \end{equation}
    
        \item Matching to SMEFT:
    
    \begin{equation}
    \begin{split}
        \mathcal L_{\mathrm{SMEFT}} \supset & \frac{(y^{ql}_{\omega_1})_{1}^*(y^{ql}_{\omega_1})_{3}}{4M_{\omega_1}^2}\left( [Q_{lq}^{(1)}]_{\ell\ell13} - [Q_{lq}^{(3)}]_{\ell\ell13} \right) + (\ell\ell11 + \ell\ell33)
    \end{split}
    \label{eq:SMEFTLagrangian_omega1_lq1}
    \end{equation}
    \end{itemize}

    \item $Q_{lequ}^{(1)}$ and $Q_{lequ}^{(3)}$
    \begin{itemize}
        \item UV Lagrangian:
    
    \begin{equation}
        -\mathcal L^{\leq 4} \supset (y^{ql}_{\omega_1})_{3}~ \omega_{1\ell}^\dagger \bar q_{3}^c i\sigma_2 l_{\ell} + (y^{eu}_{\omega_1})_{i}~ \omega_{1\ell}^\dagger \bar e_{\ell}^c u_{1} + \mathrm{h.c.} 
    \end{equation}
    
        \item Matching to SMEFT:
    
    \begin{equation}
    \begin{split}
        \mathcal L_{\mathrm{SMEFT}} \supset & \frac{(y^{eu}_{\omega_1})_{1}(y^{ql}_{\omega_1})_{3}^*}{8M_{\omega_1}^2}\left( 4[Q_{lequ}^{(1)}]_{\ell\ell31} - [Q_{lequ}^{(3)}]_{\ell\ell31} \right) \\
        & + \frac{|(y^{ql}_{\omega_1})_{3}|^2}{4M_{\omega_1}^2}\left( [Q_{lq}^{(1)}]_{\ell\ell33} - [Q_{lq}^{(3)}]_{\ell\ell33} \right) + \frac{|(y^{eu}_{\omega_1})_{1}|^2}{2M_{\omega_1}^2} [Q_{eu}]_{\ell\ell11}
    \end{split}
    \end{equation}
    \end{itemize}
    
\end{itemize}

\subsubsection*{$\Pi_7 \sim (\mathbf{3},\mathbf{2},\frac{7}{6})_S$}

\begin{itemize}
    \item $Q_{lequ}^{(1)}$ and $Q_{lequ}^{(3)}$
    \begin{itemize}
        \item UV Lagrangian:
    
    \begin{equation}
        -\mathcal L^{\leq 4} \supset (y^{lu}_{\Pi_7})_{1}~ \Pi_{7\ell}^{\dagger} i\sigma_2 \bar l_{\ell}^T u_{1} + (y^{eq}_{\Pi_7})_{3}~ \Pi_{7\ell}^{\dagger} \bar e_{\ell} q_{3} + \mathrm{h.c.}
    \end{equation}
    
        \item Matching to SMEFT:
    
    \begin{equation}
    \begin{split}
        \mathcal L_{\mathrm{SMEFT}} \supset & \frac{(y^{eq}_{\Pi_7})_{3}^*(y^{lu}_{\Pi_7})_{1}}{8M_{\Pi_7}^2}\left( 4[Q_{lequ}^{(1)}]_{\ell\ell31} + [Q_{lequ}^{(3)}]_{\ell\ell31} \right) \\
        & - \frac{|(y^{lu}_{\Pi_7})_{1}|^2}{2M_{\Pi_7}^2} [Q_{lu}]_{\ell\ell11} - \frac{|(y^{eq}_{\Pi_7})_{3}|^2}{2M_{\Pi_7}^2} [Q_{qe}]_{33\ell\ell}
    \end{split}
    \end{equation}
    \end{itemize}
\end{itemize}

\subsubsection*{$\zeta \sim (\mathbf{3},\mathbf{3},-\frac{1}{3})_S$}

\begin{itemize}
    \item $Q_{lq}^{(3)}$
    \begin{itemize}
        \item UV Lagrangian:
    
    \begin{equation}
        -\mathcal L^{\leq 4} \supset (y^{ql}_{\zeta})_{1}~ \zeta^{a\dagger}_{\ell} \bar q_{1}^c i\sigma_2\sigma^a l_{\ell} + (y^{ql}_{\zeta})_{3}~ \zeta^{a\dagger}_{\ell} \bar q_{3}^c i\sigma_2\sigma^a l_{\ell} + \mathrm{h.c.}
    \end{equation}
    
        \item Matching to SMEFT:
    
    \begin{equation}
    \begin{split}
        \mathcal L_{\mathrm{SMEFT}} \supset & \frac{(y^{ql}_{\zeta})_{1}^*(y^{ql}_{\zeta})_{3}}{4M_{\zeta}^2}\left( 3[Q_{lq}^{(1)}]_{\ell\ell13} + [Q_{lq}^{(3)}]_{\ell\ell13} \right) + (\ell\ell11 + \ell\ell33)
    \end{split}
    \end{equation}
    \end{itemize}
\end{itemize}

\subsection{Vector-like fermions}

\subsubsection*{$U \sim (\mathbf{3},\mathbf{1},\frac{2}{3})_F$}

\begin{itemize}
    \item $Q_{\phi q}^{(3)}$
    \begin{itemize}
        \item UV Lagrangian:
    
    \begin{equation}
        -\mathcal L^{\leq 4} \supset (\lambda_{U})_1~ \bar U_{R} \tilde\phi^\dagger q_{1} + (\lambda_{U})_3~ \bar U_{R} \tilde\phi^\dagger q_{3} + \mathrm{h.c.}
    \end{equation}
    
        \item Matching to SMEFT:
    
    \begin{equation}
    \begin{split}
        \mathcal L_{\mathrm{SMEFT}} \supset & \frac{(\lambda_U)_{3}(\lambda_U)_{1}^*}{4M_{U}^2}\left( [Q_{\phi q}^{(1)}]_{13} - [Q_{\phi q}^{(3)}]_{13} \right) + (11 + 33) \\
        & + \left(\frac{\hat y ^{u*}_{i1}|(\lambda_U)_1|^2}{2M_U^2} + \frac{\hat y ^{u*}_{i3}(\lambda_U)_3(\lambda_U)_1^*}{2M_U^2}\right)[Q_{u\phi}]_{1i} \\
        & + \left(\frac{\hat y ^{u*}_{i3}|(\lambda_U)_3|^2}{2M_U^2} + \frac{\hat y ^{u*}_{i1}(\lambda_U)_1(\lambda_U)_3^*}{2M_U^2}\right)[Q_{u\phi}]_{3i}
    \end{split}
    \end{equation}
    with the index $i$ running over all three generations.
    \end{itemize}
    
\end{itemize}

\subsubsection*{$D \sim (\mathbf{3},\mathbf{1},-\frac{1}{3})_F$}

\begin{itemize}
    \item $Q_{\phi q}^{(3)}$
    \begin{itemize}
        \item UV Lagrangian:
    
    \begin{equation}
        -\mathcal L^{\leq 4} \supset (\lambda_{D})_1~ \bar D_{R} \phi^\dagger q_{L1} + (\lambda_{D})_3~ \bar D_{R} \phi^\dagger q_{L3} + \mathrm{h.c.}
    \end{equation}
    
        \item Matching to SMEFT:
    
    \begin{equation}
    \begin{split}
        \mathcal L_{\mathrm{SMEFT}} \supset & -\frac{(\lambda_D)_{3}(\lambda_D)_{1}^*}{4M_{D}^2}\left( [Q_{\phi q}^{(1)}]_{13} + [Q_{\phi q}^{(3)}]_{13} \right) + (11 + 33) \\
        & + \left(\frac{\hat y ^{d*}_{i1}|(\lambda_D)_1|^2}{2M_D^2} + \frac{\hat y ^{d*}_{i3}(\lambda_D)_3(\lambda_D)_1^*}{2M_D^2}\right)[Q_{d\phi}]_{1i} \\
        & + \left(\frac{\hat y ^{d*}_{i3}|(\lambda_D)_3|^2}{2M_D^2} + \frac{\hat y ^{d*}_{i1}(\lambda_D)_1(\lambda_D)_3^*}{2M_D^2}\right)[Q_{d\phi}]_{3i}
    \end{split}
    \end{equation}
    \end{itemize}
    with the index $i$ running over all three generations.
\end{itemize}

\subsubsection*{$Q_1 \sim (\mathbf{3},\mathbf{2},\frac{1}{6})_F$}

\begin{itemize}
    \item $Q_{\phi ud}$
    \begin{itemize}
        \item UV Lagrangian:
    
    \begin{equation}
        -\mathcal L^{\leq 4} \supset (\lambda_{Q_1}^u)_1~ \bar Q_{1L} \tilde\phi u_{1} + (\lambda_{Q_1}^d)_3~ \bar Q_{1L} \phi d_{3} + \mathrm{h.c.}
        \label{eq:UVLagrangian_Q1}
    \end{equation}
    
        \item Matching to SMEFT:
    
    \begin{equation}
    \begin{split}
        \mathcal L_{\mathrm{SMEFT}} \supset & \frac{(\lambda^d_{Q_1})_{3}(\lambda^u_{Q_1})_{1}^*}{M_{Q_1}^2}[Q_{\phi ud}]_{13} - \frac{|(\lambda^u_{Q_1})_{1}|^2}{2M_{Q_1}^2}[Q_{\phi u}]_{11} + \frac{|(\lambda^d_{Q_1})_{3}|^2}{2M_{Q_1}^2}[Q_{\phi d}]_{33} \\
        & + \frac{\hat y ^{u*}_{1i}|(\lambda^u_{Q_1})_1|^2}{2M_{Q_1}^2} [Q_{u\phi}]_{i1} + \frac{\hat y ^{d*}_{3i}|(\lambda^d_{Q_1})_3|^2}{2M_{Q_1}^2}[Q_{d\phi}]_{i3}
        \label{eq:SMEFTLagrangian_Q1}
    \end{split}
    \end{equation}
    with the index $i$ running over all three generations.
    \end{itemize}
\end{itemize}

\subsubsection*{$T_1 \sim (\mathbf{3},\mathbf{3},-\frac{1}{3})_F$}

\begin{itemize}
    \item $Q_{\phi q}^{(3)}$
    \begin{itemize}
        \item UV Lagrangian:
    
    \begin{equation}
        -\mathcal L^{\leq 4} \supset \frac{1}{2}(\lambda_{T_1})_1~ \bar T_{1R}^a \phi^\dagger \sigma^a q_{1} + \frac{1}{2}(\lambda_{T_1})_3~ \bar T_{1R}^a \phi^\dagger \sigma^a q_{3} + \mathrm{h.c.}
    \end{equation}
    
        \item Matching to SMEFT:
    
    \begin{equation}
    \begin{split}
        \mathcal L_{\mathrm{SMEFT}} \supset & -\frac{(\lambda_{T_1})_{3}(\lambda_{T_1})_{1}^*}{16M_{T_1}^2}\left( 3[Q_{\phi q}^{(1)}]_{13} - [Q_{\phi q}^{(3)}]_{13} \right) + (11 + 33) \\
        & + \left(\frac{\hat y ^{d*}_{i1}|(\lambda_{T_1})_1|^2}{8M_{T_1}^2} + \frac{\hat y ^{d*}_{i3}(\lambda_{T_1})_3(\lambda_{T_1})_1^*}{8M_{T_1}^2}\right)[Q_{d\phi}]_{1i} \\
        & + \left(\frac{\hat y ^{d*}_{i3}|(\lambda_{T_1})_3|^2}{8M_{T_1}^2} + \frac{\hat y ^{d*}_{i1}(\lambda_{T_1})_1(\lambda_{T_1})_3^*}{8M_{T_1}^2}\right)[Q_{d\phi}]_{3i} \\
        & + \left(\frac{\hat y ^{u*}_{i1}|(\lambda_{T_1})_1|^2}{4M_{T_1}^2} + \frac{\hat y ^{u*}_{i3}(\lambda_{T_1})_3(\lambda_{T_1})_1^*}{4M_{T_1}^2}\right)[Q_{u\phi}]_{1i} \\
        & + \left(\frac{\hat y ^{u*}_{i3}|(\lambda_{T_1})_3|^2}{4M_{T_1}^2} + \frac{\hat y ^{u*}_{i1}(\lambda_{T_1})_1(\lambda_{T_1})_3^*}{4M_{T_1}^2}\right)[Q_{u\phi}]_{3i}
    \end{split}
    \end{equation}
    with the index $i$ running over all three generations.
    \end{itemize}
\end{itemize}

\subsubsection*{$T_2 \sim (\mathbf{3},\mathbf{3},\frac{2}{3})_F$}

\begin{itemize}
    \item $Q_{\phi q}^{(3)}$
    \begin{itemize}
        \item UV Lagrangian:
    
    \begin{equation}
        -\mathcal L^{\leq 4} \supset \frac{1}{2}(\lambda_{T_2})_1~ \bar T_{2R}^a \tilde\phi^\dagger \sigma^a q_{1} + \frac{1}{2}(\lambda_{T_2})_3~ \bar T_{2R}^a \tilde\phi^\dagger \sigma^a q_{3} + \mathrm{h.c.}
    \end{equation}
    
        \item Matching to SMEFT:
    
    \begin{equation}
    \begin{split}
        \mathcal L_{\mathrm{SMEFT}} \supset & \frac{(\lambda_{T_2})_{3}(\lambda_{T_2})_{1}^*}{16M_{T_2}^2}\left( 3[Q_{\phi q}^{(1)}]_{13} + [Q_{\phi q}^{(3)}]_{13} \right) + (11 + 33) \\
        & + \left(\frac{\hat y ^{d*}_{i1}|(\lambda_{T_1})_1|^2}{4M_{T_2}^2} + \frac{\hat y ^{d*}_{i3}(\lambda_{T_1})_3(\lambda_{T_1})_1^*}{4M_{T_2}^2}\right)[Q_{d\phi}]_{1i} \\
        & + \left(\frac{\hat y ^{d*}_{i3}|(\lambda_{T_1})_3|^2}{4M_{T_2}^2} + \frac{\hat y ^{d*}_{i3}(\lambda_{T_1})_1(\lambda_{T_1})_3^*}{4M_{T_2}^2}\right)[Q_{d\phi}]_{3i} \\
        & + \left(\frac{\hat y ^{u*}_{i1}|(\lambda_{T_1})_1|^2}{8M_{T_2}^2} + \frac{\hat y ^{u*}_{i3}(\lambda_{T_1})_3(\lambda_{T_1})_1^*}{8M_{T_2}^2}\right)[Q_{u\phi}]_{1i} \\
        & + \left(\frac{\hat y ^{u*}_{i3}|(\lambda_{T_1})_3|^2}{8M_{T_2}^2} + \frac{\hat y ^{u*}_{i3}(\lambda_{T_1})_1(\lambda_{T_1})_3^*}{8M_{T_2}^2}\right)[Q_{u\phi}]_{3i}
    \end{split}
    \end{equation}
    with the index $i$ running over all three generations.
    \end{itemize}
\end{itemize}

\subsection{Vectors}

\subsubsection*{$\mathcal B_1 \sim (\mathbf{1},\mathbf{1},1)_V$}
\begin{itemize}
    \item $Q_{\phi ud}$
    \begin{itemize}
        \item UV Lagrangian:
    
    \begin{equation}
        -\mathcal L^{\leq 4} \supset (g_{\mathcal B_1}^{du})_{31}~ \mathcal B_1^{\mu\dagger} \bar d_{3} \gamma_\mu u_{1} + (g_{\mathcal B_1}^\phi)~ \mathcal B_1^{\mu\dagger} iD_\mu \phi^T i\sigma_2\phi + \mathrm{h.c.}
    \end{equation}
    
        \item Matching to SMEFT:
    
    \begin{equation}
    \begin{split}
        \mathcal L_{\mathrm{SMEFT}} \supset & -\frac{(g_{\mathcal B_1}^\phi)(g_{\mathcal B_1}^{du})^*_{31}}{M_{\mathcal B_1}^2}[Q_{\phi ud}]_{13} - \frac{|(g_{\mathcal B_1}^{du})_{31}|^2}{3M_{\mathcal B_1}^2}\left( [Q_{ud}^{(1)}]_{1133} + 6[Q_{ud}^{(8)}]_{1133} \right) \\
        & - \frac{\hat y^{u*}_{ji}|g_{\mathcal B_1}^\phi|^2}{2M_{\mathcal B_1}^2}[Q_{u\phi}]_{ij} - \frac{\hat y^{d*}_{ji}|g_{\mathcal B_1}^\phi|^2}{2M_{\mathcal B_1}^2}[Q_{d\phi}]_{ij} - \frac{\hat y^{e*}_{ji}|g_{\mathcal B_1}^\phi|^2}{2M_{\mathcal B_1}^2}[Q_{e\phi}]_{ij} \\
        & + \frac{|g_{\mathcal B_1}^\phi|^2}{M_{\mathcal B_1}^2}Q_{\phi D} - \frac{|g_{\mathcal B_1}^\phi|^2}{2M_{\mathcal B_1}^2}Q_{\phi \Box} - \frac{2\hat\lambda_\phi|g_{\mathcal B_1}^\phi|^2}{M_{\mathcal B_1}^2}Q_{\phi} + \frac{\mu_\phi^2|g_{\mathcal B_1}^\phi|^2}{M_{\mathcal B_1}^2}Q_{\phi4}
    \end{split}
    \end{equation}
        with indices $i,j$ running over all three generations, $\hat\lambda_\phi = \lambda_\phi - C_{\phi4}$, $\lambda_\phi$ and $\mu_\phi$ being the parameters of the Higgs potential and $C_{\phi4}$ being the coefficient of the operator $Q_{\phi4}$ given in the last term.
    \end{itemize}
    
\end{itemize}

\subsubsection*{$\mathcal W \sim (\mathbf{1},\mathbf{3},0)_V$}
\begin{itemize}
    \item $Q_{lq}^{(3)}$
    \begin{itemize}
        \item UV Lagrangian:
    
    \begin{equation}
        -\mathcal L^{\leq 4} \supset \frac{1}{2}(g_{\mathcal W}^{q})_{13}~ \mathcal W^{\mu a} \bar q_{1} \sigma^a\gamma_\mu q_{3} + \frac{1}{2}(g_{\mathcal W}^l)~ \mathcal W^{\mu a} \bar l_{\ell} \sigma^a\gamma_\mu l_{\ell}
    \end{equation}
    
        \item Matching to SMEFT:
    
    \begin{equation}
    \begin{split}
        \mathcal L_{\mathrm{SMEFT}} \supset & -\frac{(g_{\mathcal W}^q)_{13}(g_{\mathcal W}^{l})}{4M_{\mathcal W}^2}[Q_{lq}^{(3)}]_{\ell\ell13} - \frac{(g_{\mathcal W}^q)_{13}^2}{8M_{\mathcal W}^2}[Q_{qq}^{(3)}]_{1313} \\
        & - \frac{(g_{\mathcal W}^l)^2}{4M_{\mathcal W}^2}[Q_{ll}]_{\ell\ell'\ell'\ell} + \frac{(g_{\mathcal W}^l)^2}{8M_{\mathcal W}^2}[Q_{ll}]_{\ell\ell\ell'\ell'}
    \end{split}
    \end{equation}
    
    \end{itemize}

    \item $Q_{\phi q}^{(3)}$
    \begin{itemize}
        \item UV Lagrangian:
    
    \begin{equation}
        \mathcal L \supset \frac{1}{2}(g_{\mathcal W}^{q})_{13}~ \mathcal W^{\mu a} \bar q_{1} \sigma^a\gamma_\mu q_{3} + \left\{\frac{1}{2}(g_{\mathcal W}^\phi)~ \mathcal W^{\mu a} \phi^\dagger \sigma^a i D_\mu \phi + \mathrm{h.c.}\right\}
    \end{equation}
    
        \item Matching to SMEFT:
    
    \begin{equation}
    \begin{split}
        \mathcal L_{\mathrm{SMEFT}} \supset & -\frac{\mathrm{Re}(g_{\mathcal W}^\phi)(g_{\mathcal W}^q)_{13}}{4M_{\mathcal W}^2}[Q_{\phi q}^{(3)}]_{13} - \frac{(g_{\mathcal W}^q)_{13}^2}{8M_{\mathcal W}^2}[Q_{qq}^{(3)}]_{1313} \\
        & - \left(\frac{\hat y^{u*}_{ji}|g_{\mathcal W}^\phi|^2}{4M_{\mathcal W}^2} - \frac{i\hat y^{u*}_{ji}\mathrm{Im}\left((g_{\mathcal W}^\phi)^2\right)}{8M_{\mathcal W}^2} \right)[Q_{u\phi}]_{ij} \\
        & - \left(\frac{\hat y^{d*}_{ji}|g_{\mathcal W}^\phi|^2}{4M_{\mathcal W}^2} + \frac{i\hat y^{d*}_{ji}\mathrm{Im}\left((g_{\mathcal W}^\phi)^2\right)}{8M_{\mathcal W}^2} \right)[Q_{d\phi}]_{ij} \\
        & - \left(\frac{\hat y^{e*}_{ji}|g_{\mathcal W}^\phi|^2}{4M_{\mathcal W}^2} + \frac{i\hat y^{e*}_{ji}\mathrm{Im}\left((g_{\mathcal W}^\phi)^2\right)}{8M_{\mathcal W}^2} \right)[Q_{e\phi}]_{ij} \\
        & - \frac{i\hat y^{u*}_{i3}\mathrm{Im}(g_{\mathcal W}^\phi) (g_{\mathcal W}^q)_{13}}{4M_{\mathcal W}^2}[Q_{u\phi}]_{1i} + \frac{i\hat y^{d*}_{i3} \mathrm{Im}(g_{\mathcal W}^\phi) (g_{\mathcal W}^q)_{13}}{4M_{\mathcal W}^2}[Q_{d\phi}]_{1i} \\
        & - \left(\frac{\mathrm{Re}((g_{\mathcal W}^\phi)^2)}{4M_{\mathcal W}^2} - \frac{|g_{\mathcal W}^\phi|^2}{4M_{\mathcal W}^2}\right)Q_{\phi D} - \left(\frac{\mathrm{Re}((g_{\mathcal W}^\phi)^2)}{8M_{\mathcal W}^2} + \frac{|g_{\mathcal W}^\phi|^2}{4M_{\mathcal W}^2}\right)Q_{\phi \Box} \\ 
        & - \frac{\hat\lambda_\phi|g_{\mathcal W}^\phi|^2}{M_{\mathcal W}^2}Q_{\phi} + \frac{\hat\mu_\phi^2|g_{\mathcal W}^\phi|^2}{2M_{\mathcal W}^2}Q_{\phi4}
    \end{split}
    \end{equation}
        with indices $i,j$ running over all three generations, $\hat\lambda_\phi = \lambda_\phi - C_{\phi4}$, $\lambda_\phi$ and $\mu_\phi$ being the parameters of the Higgs potential and $C_{\phi4}$ being the coefficient of the operator $Q_{\phi4}$ given in the last term.
    \end{itemize}
    
\end{itemize}

\subsubsection*{$\mathcal U_2 \sim (\mathbf{3},\mathbf{1},\frac{2}{3})_V$}
\begin{itemize}
    \item $Q_{lq}^{(3)}$
    \begin{itemize}
        \item UV Lagrangian:
    
    \begin{equation}
        -\mathcal L^{\leq 4} \supset (g_{\mathcal U_2}^{lq})_{1}~ \mathcal U_{2\ell}^{\mu\dagger} \bar l_{\ell} \gamma_\mu q_{1} + (g_{\mathcal U_2}^{lq})_{3}~ \mathcal U_{2\ell}^{\mu\dagger} \bar l_{\ell} \gamma_\mu q_{3}  + \mathrm{h.c.}
    \end{equation}
    
        \item Matching to SMEFT:
    
    \begin{equation}
    \begin{split}
        \mathcal L_{\mathrm{SMEFT}} \supset & -\frac{(g^{lq}_{\mathcal U_2})_{1}^*(g^{lq}_{\mathcal U_2})_{3}}{2M_{\mathcal U_2}^2}\left( [Q_{lq}^{(1)}]_{\ell\ell13} + [Q_{lq}^{(3)}]_{\ell\ell13} \right) + (\ell\ell11 + \ell\ell33)
    \end{split}
    \end{equation}
    \end{itemize}
    
    \item $Q_{ledq}$
    
    \begin{itemize}
        \item UV Lagrangian:
    
    \begin{equation}
        -\mathcal L^{\leq 4} \supset (g_{\mathcal U_2}^{lq})_{1}~ \mathcal U_{2\ell}^{\mu\dagger} \bar l_{\ell} \gamma_\mu q_{1} + (g_{\mathcal U_2}^{ed})_{3}~ \mathcal U_{2\ell}^{\mu\dagger} \bar e_{\ell} \gamma_\mu d_{3}  + \mathrm{h.c.}
    \end{equation}
    
        \item Matching to SMEFT:
    
    \begin{equation}
    \begin{split}
        \mathcal L_{\mathrm{SMEFT}} \supset & \frac{2(g^{lq}_{\mathcal U_2})_{1}(g^{ed}_{\mathcal U_2})_{3}}{M_{\mathcal U_2}^2}[Q_{ledq}]_{\ell\ell31} -\frac{|(g^{lq}_{\mathcal U_2})_{1}|^2}{2M_{\mathcal U_2}^2}\left( [Q_{lq}^{(1)}]_{\ell\ell11} + [Q_{lq}^{(3)}]_{\ell\ell11} \right)\\
        & -\frac{|(g^{ed}_{\mathcal U_2})_{3}|^2}{M_{\mathcal U_2}^2}[Q_{ed}]_{\ell\ell33}
    \end{split}
    \end{equation}
    \end{itemize}

\end{itemize}

\subsubsection*{$\mathcal Q_5 \sim (\mathbf{3},\mathbf{2},-\frac{5}{6})_V$}
\begin{itemize}
    \item $Q_{ledq}$
    \begin{itemize}
        \item UV Lagrangian:
    
    \begin{equation}
        -\mathcal L^{\leq 4} \supset (g_{\mathcal Q_5}^{eq})_{1}~ \mathcal Q_{5\ell}^{\mu \dagger} \bar e_{\ell}^c \gamma_\mu q_{1} + (g_{\mathcal Q_5}^{dl})_{3}~ \mathcal Q_{5\ell}^{\mu \dagger} \bar d_{3}^c \gamma_\mu l_{\ell}  + \mathrm{h.c.}
    \end{equation}
    
        \item Matching to SMEFT:
    
    \begin{equation}
    \begin{split}
        \mathcal L_{\mathrm{SMEFT}} \supset & -\frac{2(g^{eq}_{\mathcal Q_5})_{1}(g^{dl}_{\mathcal Q_5})_{3}^*}{M_{\mathcal Q_5}^2} [Q_{ledq}]_{\ell\ell31} + \frac{|(g^{eq}_{\mathcal Q_5})_{\ell1}|^2}{M_{\mathcal Q_5}^2} [Q_{qe}]_{11\ell\ell}
    \end{split}
    \end{equation}
    \end{itemize}
\end{itemize}

\subsubsection*{$\mathcal X \sim (\mathbf{3},\mathbf{3},\frac{2}{3})_V$}
\begin{itemize}
    \item $Q_{lq}^{(3)}$
    \begin{itemize}
        \item UV Lagrangian:
    
    \begin{equation}
        -\mathcal L^{\leq 4} \supset \frac{1}{2}(g_{\mathcal X})_{1}~ \mathcal X^{\mu a\dagger}_\ell \bar l_{\ell} \gamma_\mu \sigma^a q_{1} + \frac{1}{2}(g_{\mathcal X})_{3}~ \mathcal X^{\mu a\dagger}_\ell \bar l_{\ell} \gamma_\mu \sigma^a q_{3}  + \mathrm{h.c.}
    \end{equation}
    
        \item Matching to SMEFT:
    
    \begin{equation}
    \begin{split}
        \mathcal L_{\mathrm{SMEFT}} \supset & -\frac{(g_{\mathcal X})_{1}^*(g_{\mathcal X})_{3}}{8M_{\mathcal X}^2}\left( 3[Q_{lq}^{(1)}]_{\ell\ell13} - [Q_{lq}^{(3)}]_{\ell\ell13} \right) + (\ell\ell11 + \ell\ell33)
    \end{split}
    \end{equation}
    \end{itemize}

\end{itemize}

\bibliography{references}
\bibliographystyle{JHEP}

\end{document}